\documentclass[pra,aps,amsmath,amssymb,amsfonts,twocolumn,nofootinbib,longbibliography,superscriptaddress]{revtex4-2}
\usepackage{amssymb}
\usepackage{bm,mathrsfs}
\usepackage{graphicx}
\usepackage{subfigure}
\usepackage{epsfig}
\usepackage{amsmath,bbm}
\usepackage{amsfonts,amssymb}
\usepackage{times}
\usepackage{verbatim}
\usepackage{diagbox}
\usepackage[sort&compress]{natbib}
\usepackage{amsmath}
\usepackage[colorlinks=true,citecolor=blue,linkcolor=blue,urlcolor=blue]{hyperref}
\usepackage[usenames]{color}

\usepackage{float}
\begin{document}

\preprint{APS/123-QED}

\title{Localization and Dynamics in One-dimensional and Two-dimensional Gaussian Disordered Quantum Spin Systems}

\author{Dongyan Guo}
\affiliation{Center for Quantum Sciences and School of Physics, Northeast Normal University, Changchun 130024, China}

\author{Taotao Hu}
\email{hutt262@nenu.edu.cn}
\affiliation{Center for Quantum Sciences and School of Physics, Northeast Normal University, Changchun 130024, China}

\author{Jiameng Hong}
\affiliation{Center for Quantum Sciences and School of Physics, Northeast Normal University, Changchun 130024, China}

\begin{abstract}
This paper introduces a new form of disorder — Gaussian disorder. Gaussian disorder involves two parameters: the expected value $\mu$ and the standard deviation $\sigma$. Studying Gaussian disorder can enhance our understanding of how many-body localization (MBL) transition is influenced by the central value and the breadth of disorder. We apply Gaussian disorder to a one-dimensional spin-1/2 chain with nearest-neighbor and next-nearest-neighbor couplings. By analyzing the properties of eigenstates and dynamical characteristics of the disordered system, we reveal the features of Gaussian disorder. It shows that Gaussian disorder can drive localization transitions in many-body systems. The crucial parameter governing this process is the standard deviation. As the standard deviation increases, the system shows a transition from the ergodic phase to the localized phase, while the other parameter of Gaussian disorder, the expected value, has no effect to the transition. Our findings indicate that the critical factor influencing MBL transition is the breadth of disorder rather than its central value. Then, we apply Gaussian disorder to two different system-environment models (hereafter referred to as the sye-env models, consisting of ladder configuration and staggered configuration) and analyze the mutual influence between the system and the environment when they are coupled. The steady-state staggered magnetization is used to indicate the eventual localization properties of the system and environment after a quench from the initial state. We extensively explore the impacts brought about by system disorder, environment disorder, initial state, and the strength of the sys-env interaction. Finally, for theses two different models, we consider the system and environment as a whole, studying how the localization properties of the entire system are influenced by the disorder strength of each chain and the interaction strength between the two chains. It is worth noting that in the ladder configuration, we observe a strong dependence on the initial state, which can lead to either the loss or retention of information, whereas no such dependence on the initial state is found in the staggered configuration. We emphasize that when using dynamic indicators to analyze system characteristics, it is essential to consider the influence of different initial states, as this may lead to significant differences under certain structural conditions, and not all initial states can accurately reflect the localization properties of the many-body system.

\end{abstract}

\maketitle

\section{\label{sec:level1}Introduction}
The phenomenon of localization was first proposed by Anderson in 1958\cite{anderson1958absence}, and subsequently, researchers expanded upon his work, leading to the development of many-body localization. The study of many-body quantum systems which deviate from the eigenstate thermalization hypothesis (ETH)\cite{anderson1958absence,PhysRevE.50.888,deutsch1991quantum,doi:10.1080/00018732.2016.1198134} is a significant frontier in modern condensed matter physics. With continuous and in-depth research, the fundamental characteristics of the MBL have been discovered.

Theoretical studies of MBL typically consider a uniform, uncorrelated random potential as the source of disorder in the system. It is now known that strong disorder can drive closed ergodic systems into localization\cite{serbyn2013local,huse2014phenomenology,schreiber2015observation,oganesyan2007localization,luitz2015many}. When the system is in the two different phases (localized and ergodic phases), its eigenstates exhibit distinct characteristics\cite{serbyn2013local,deutsch1991quantum,pal2010many}. In recent years, the dynamical characteristics of many-body systems have also attracted increasing attention. The dynamics of most physical quantities generally show deviation from their initial values over time, eventually reaching stable values\cite{thomson2023localization,wu2016understanding,chiew2023stability}. Moreover, in the localized phase, the deviation of stable values from their initial values is smaller compared to that in the ergodic phase.

Moreover, in practical situations, quantum systems are not completely isolated but are coupled to external environments to some extent. The localization properties of the systems can be influenced by the environments, making it meaningful to consider systems that are coupled to their environments. When a system is coupled to a dissipative environment, it is expected to thermalize over long time scales\cite{levi2016robustness,fischer2016dynamics,medvedyeva2016influence,wei2018exploring}. However, if the number of degrees of freedom of the environment is comparable to that of the system, under certain conditions, the significant backaction and proximity effects can prevent quantum systems from thermalization\cite{chiew2023stability}.

In this paper, we introduce a new form of disorder: Gaussian disorder, and apply it to a one-dimensional chain. We calculate physical quantities commonly used in the study of MBL transitions, which reflect the properties of eigenstates and dynamical characteristics of the disordered system. Gaussian disorder involves two parameters: the standard deviation and the expected value. As the standard deviation increases, the system shows a transition from the ergodic phase to the localized phase, while the expected value does not affect the phase transition. The results indicate that the intrinsic mechanism triggering localization transitions in many-body systems is the breadth of disorder rather than its central value.

Next, we apply Gaussian disorder to the sys-env models, placing the system and environment in two different forms of disordered external fields. The mutual influence between the system and environment is separately analyzed in two different sys-env models (ladder configuration and staggered configuration). These two models simulate scenarios with simple and complex interaction strengths between the system and the environment. We use the steady-state staggered magnetization as an indicator to illustrate how the system and environment preserve their initial information when coupled together. The eventual localization properties of them are influenced by the system disorder, environmental disorder, sys-env interaction strength, and initial state of the whole system. The results indicate that when the interaction strength between the system and environment is weak, the two chains behave nearly independently, and both can retain their initial information without being influenced by the other. As the interaction strength increases, the localization properties of the system (environment) gradually become influenced by the disorder of the environment (system), but the key factor is still the inherent disorder. When the system and environment are strongly coupled, there is a high correlation between them. The impact between the system and the environment has increased, especially for initially ergodic systems and environments, which are more susceptible to each other's influence. Additionally, it is worth noting that in this case, for the ladder configuration, the eventual localization properties of them are affected by the initial state, which is absent in the staggered configuration.

  Finally, we treat the system and environment as a whole to study the localization of the entire system. Unlike previous studies\cite{baygan2015many,bouillot2011statics,sun2020characterizing,yamaguchi2014field}, in our research, each chain is driven by different forms of disordered external fields. The localization properties of the entire system is influenced by the disorder of each chain. Even if the disorder of one chain is weak, the entire system can exhibit MBL transition if the disorder of the other chain is sufficiently strong. Then we analyzed how the dynamical characteristics of the entire system change with \( J_{int} \) ( interaction strength between the system and the environment) for different initial states. We find that in the ladder configuration, different initial states yield different dynamical results (especially at strong \( J_{int} \)), while in the staggered configuration, consistent results are obtained. This is due to the variations in energy density caused by the initial states, which are more pronounced in the ladder configuration and weaker in the staggered configuration. A related issue is the existence of MBL mobility edges\cite{luitz2015many,kjall2014many,mondragon2015many,xu2020dynamical,wei2019investigating,wei2019investigating,devakul2015early,geissler2020mobility,de2016absence,wei2020characterization,yousefjani2023mobility}, where states at different energy densities may exhibit different localization properties. This indicates that we cannot directly use the results from dynamic indicators to explain the localization properties of the system. By comparing with the localization properties reflected by the system's eigenstates, we conclude that in certain models, not all dynamical results from initial states can reflect the localization properties of the system. Therefore, it is necessary to select appropriate initial states, if we want to accurately reflect the system's localization properties through dynamic indicators.

  The paper is organized as follows. Section \ref{sec2} introduces the Gaussian disorder, one-dimensional model, and sys-env models in different interaction configurations. In Section \ref{sec3}, we calculate the physical quantities that reflect the properties of eigenstates and dynamical characteristics of the one-dimensional chain to explain the features of Gaussian disorder. In Section \ref{sec4}, we use the steady-state staggered magnetization as an indicator to investigate how the eventual localization properties of the system and environment in two different configurations are influenced by system disorder, environment disorder, the interaction strength between the system and the environment, and initial state. In Section \ref{sec5}, we separately consider the two sys-env models as a whole and study how disorder on both sides, as well as the interaction strength between the two chains, influence the localization properties of the entire system. We summarize our findings in Section \ref{sec6}.

\section{\label{sec2}models}
\subsection{One-dimensional spin-1/2 model}

	A commonly used model in many-body localization studies is the one-dimensional spin-1/2 Heisenberg chain\cite{agarwal2015anomalous,luitz2015many,bera2015many,bera2017density,herviou2019multiscale,colmenarez2019statistics}. The first model in this paper is a single chain based on the Heisenberg model with nearest and next-nearest neighbor couplings. The system's Hamiltonian can be expressed as:

 \begin{align}
	H=\sum_{i=1}^{L}{ h}_{i}S_{i}^{z}+\sum_{i=1}^{L-1}{J}_{1}\vec{S}_{i}\cdot\vec{S}_{i+1}+\sum_{i=1}^{L-2}{J}_{2}\vec{S}_{i}\cdot\vec{S}_{i+2},
  \label{eq1}
	\end{align}

 Here, \( J_1 \) and \( J_2 \) respectively denote the strength of nearest-neighbor and next-nearest-neighbor couplings, and \( L \) is the chain length. \( h_i \) represents the local external field applied at the \(i\)-th site. \( h_i \) follows a Gaussian distribution:

	\begin{align}
  P(h_{i})=\frac{1}{{\sqrt{2\pi}\sigma}}\exp[-(h_{i}-\mu)^{2}/(2\sigma^{2})].
  \label{eq2}
	\end{align}

\begin{figure}[t]
		\centering
		\includegraphics[width=1\columnwidth]{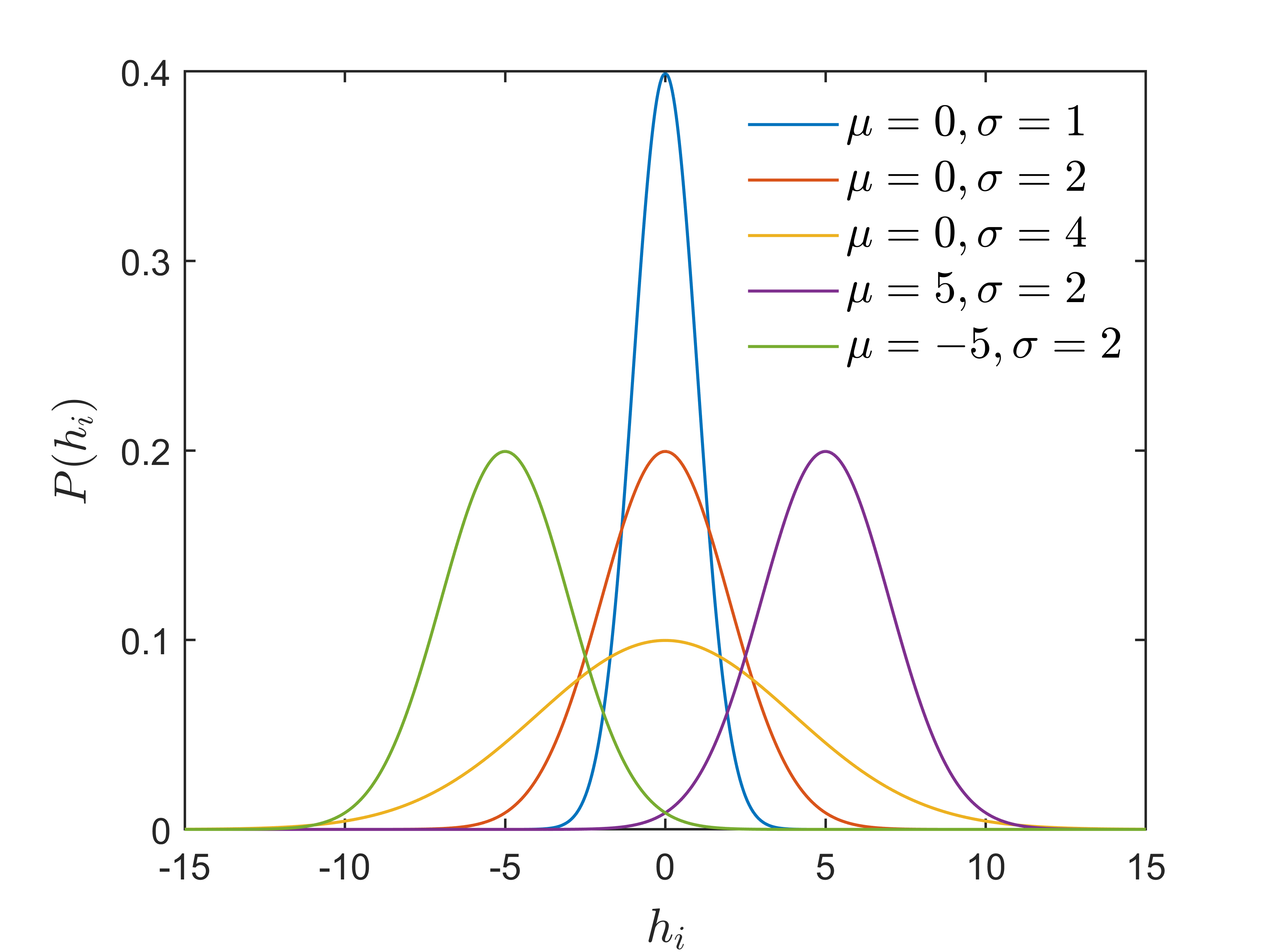}
		\caption{The probability distribution graph of the local potential \( h_i \) at arbitrary sites under Gaussian disorder, with different expected values and standard deviations. A larger standard deviation $ \sigma $ results in a broader probability distribution of disorder, while a larger expected value $ \mu $ leads to a greater central value of disorder.}
		\label{fig1}
	\end{figure}

Here, $ \mu $ is the expected value, $ \sigma $ is the standard deviation. The probability distribution of \( h_i \) depends on two parameters of the Gaussian distribution: $ \mu $ and $ \sigma $ (where \( h_i \) is a value randomly sampled from a set of 1000 numbers generated according to the specified $ \mu $ and $ \sigma $). $ \mu $ and $ \sigma $ can be respectively used to describe the central value and the breadth of the external field. Fig.\ref{fig1} shows the probability distribution of the local external field at arbitrary points for different expected values and standard deviations.

  Gaussian disorder differs from the conventional form of disorder extensively studied in the past: random distribution within $\left[-h,h\right]$, with \(h \) denoting the disorder strength (hereafter referred to as conventional disorder). In the conventional disorder, random numbers within $\left[-h,h\right]$ are chosen completely randomly, implying that the selection of any value within this range is equally probable, with a central value at 0. For the Gaussian disorder, the disordered external field acting on site \( i \) is not uniformly distributed. From Fig.\ref{fig1}, it can be seen that when the standard deviation $ \sigma $ is fixed, there is a greater probability for $\left|h_{i}-{\mu}\right|$ to take smaller values. As the standard deviation increases, $\left|h_{{i}}-{\mu}\right|$ has a greater probability of taking larger values. The expected value $ \mu $ acts as the central value of the distribution, governing the positions of disorder. In the study of Gaussian disorder, the disorder is described using the standard deviation $ \sigma $ and the expected value $ \mu $. This form of disorder facilitates our investigation into how the central value and the breadth of disorder affect the MBL transition.

  This paper concludes that as the breadth of disorder increases, the system shows a transition from the ergodic phase to the localized phase. Furthermore, the central value of the disorder does not affect the occurrence of the MBL transition. This conclusion will be demonstrated in Sec.\ref{sec3}.

 \subsection{\label{ladder mod}Sys-env model in ladder configuration} 

\begin{figure}[t]	
  \centering
		\includegraphics[width=1\columnwidth]{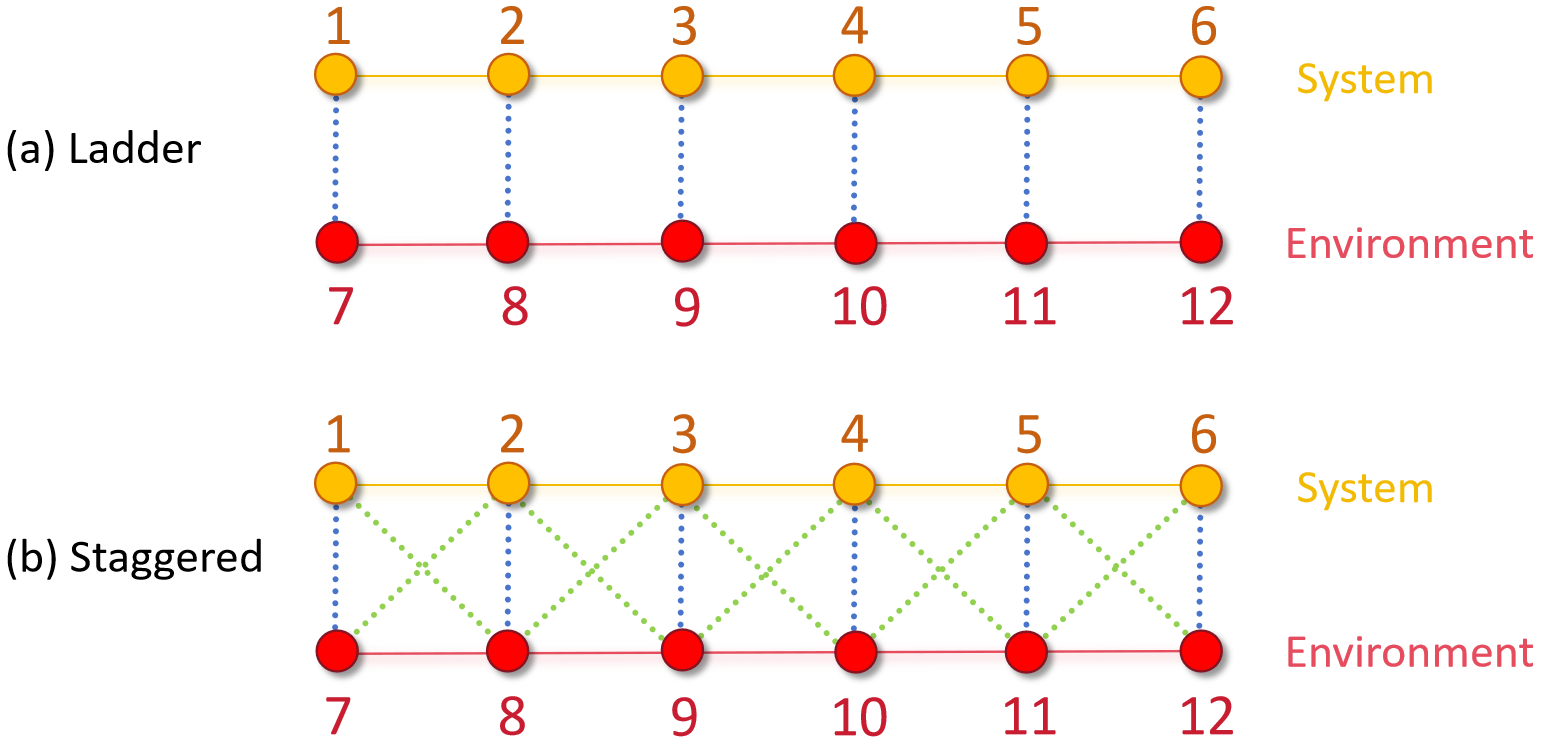}
		\caption{We consider two types of sys-env coupling configurations. Yellow (Red) chain -- system (environment), blue (green) dashed lines -- nearest-neighbor (next-nearest-neighbor) coupling between system and environment. (a) the system is attached to the environment in a ladder configuration. (b) the system is attached to the environment in a staggered configuration.}
		\label{fig2}
	\end{figure}

	The total Hamiltonian of the system and its environment can be written as:

	\begin{align}
	H_{t o t}=H_{sys}+H_{e n v}+H_{int},
   \label{eq3}
       \end{align}

where \( H_{sys} \), \( H_{env} \) and \( H_{int} \) are the system, environment and
interaction Hamiltonians, respectively. In this model, the system and the environment are spin-1/2 chains of length $l=6$ (forming a total of $L = 2l = 12$ spins), with both nearest-neighbor and next-nearest-neighbor couplings considered. They are coupled according to the ladder configuration shown in Fig.\ref{fig2}\textcolor{blue}{(a)}.

\( H_{sys} \), \( H_{env} \) and \( H_{int} \) in this case are written as:

	\begin{align}
	{H}_{s y s}&=\sum_{i=1}^{l}{h}_{i}^{(s)}{{S}}_{i}^{z}+\sum_{i=1}^{l-1}{}J_{{s}1}{\vec{S}}_{i}\cdot{\vec{S}}_{i+1}
  \nonumber\\&+\sum_{i=1}^{l-2}{}J_{{s}2}{\vec{S}}_{i}\cdot{\vec{S}}_{i+2},
   \label{eq4}
       \end{align}
	\begin{align}
	{H}_{env}&=\sum_{i=l+1}^{2l}{h}_{i}^{(e)}{{S}}_{i}^{z}+\sum_{i=l+1}^{2l-1}{}J_{{s}1}{\vec{S}}_{i}\cdot{\vec{S}}_{i+1}
  \nonumber\\&+\sum_{i=l+1}^{2l-2}{}J_{{s}2}{\vec{S}}_{i}\cdot{\vec{S}}_{i+2},
   \label{eq5}
       \end{align}

	\begin{align}	{H}_{int}&=\sum_{{i=1}}^{l}J_{int}{\vec{S}}_{i}\cdot{\vec{S}}_{i+l}.
   \label{eq6}
       \end{align}
       
In this model, ${h}_{i}^{(s)}$ and ${h}_{i}^{(e)}$ respectively represent external fields acting on the spin \( z \)-direction at each site of the system and the environment. Here, ${h}_{i}^{(s)}$ denotes conventional disorder, while ${h}_{i}^{(e)}$ denotes Gaussian disorder. We consider \( J_{s_{1}} \) (\( J_{e_{1}} \)) and \( J_{s_{2}} \) (\( J_{e_{2}} \)) respectively as the coupling strength between nearest-neighbor and next-nearest-neighbor sites in the system (environment), with \( J_{int} \) as the nearest-neighbor coupling strength between the system and the environment. We set $ J_{s_{1}} = J_{s_{2}}= J_{e_{1}} = J_{e_{1}}=J=1 $.

 \subsection{\label{stragged mod}Sys-env model in staggered configuration} 
  
   The coupling between the system and the environment in this model is more complex. They are coupled as depicted in Fig.\ref{fig2}\textcolor{blue}{(b)}, involving both nearest-neighbor and next-nearest-neighbor couplings between the system and the environment.

   \( H_{tot} \), \( H_{sys} \) and \( H_{env} \) are still represented by Eq.\ref{eq3}, \ref{eq4}, \ref{eq5}. \( H_{int} \) in this case is written as:

	\begin{align}	{H}_{{int}}&=\sum_{i=1}^{l}J_{{int_{1}}}\vec{S}_{i}\cdot\vec{S}_{i+l}\nonumber+J_{{int_{2}}}(\sum_{i=1}^{l-1}\vec{S}_{i}\cdot\vec{S}_{i+l+1} \nonumber\\&+\sum_{i=2}^{l}\vec{S}_{i}\cdot\vec{S}_{i+l-1}).
   \label{eq7}
       \end{align}

We consider \( J_{int_{1}} \) (\( J_{int_{2}} \)) as the nearest-neighbor (next-nearest-neighbor) interaction strength between the two chains. We set $ J_{int_{1}} = J_{int_{2}}= J_{int} $.

\section{\label{sec3}Many-body localization of a spin-1/2 single chain with Gaussian disorder}

In this section, our calculations are based on the one-dimensional spin-1/2 chain with Gaussian disorder. Some commonly used physical quantities in studying MBL are computed to illustrate the features of Gaussian disorder.

	\begin{figure}[t]	
  \centering
		\includegraphics[width=1\columnwidth]{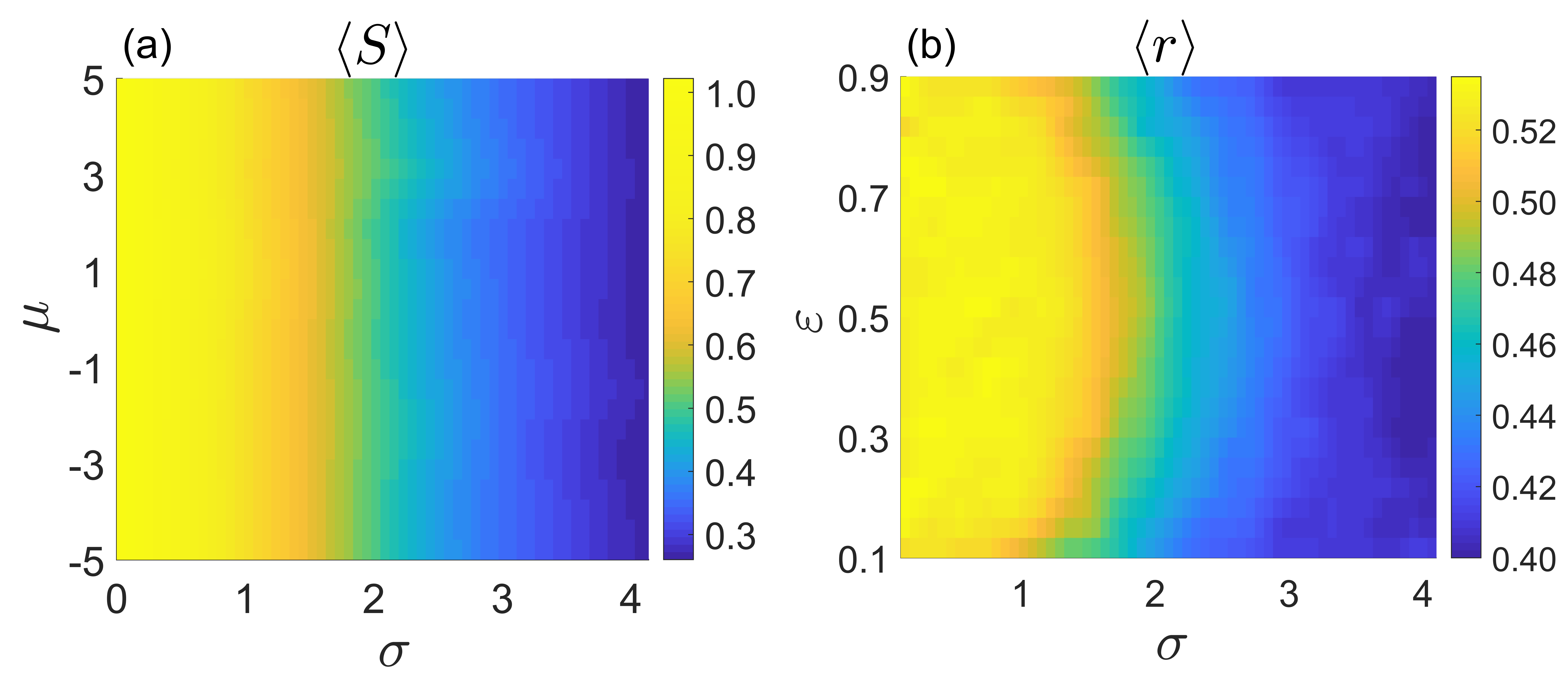}
		\caption{Consider a one-dimensional spin-1/2 chain for $L = 14$. (a) the mean entanglement entropy $\langle S\rangle$ as a function of the expected value $ \mu $ and the standard deviation $ \sigma $ of Gaussian disorder. Selecting 30 states near an energy density of $\varepsilon\approx0.5$ as samples. (b) the mean gap ratio $\langle r\rangle$ as a function of the energy density $\varepsilon$ and the standard deviation $ \sigma $ of Gaussian disorder. The “D-shape” structure in the phase diagram exhibits the well-known mobility edge. Each point is averaged 3000 disorder realisations}
		\label{fig3}
	\end{figure}

The first physical quantity to be computed is the von Neumann entanglement entropy :
	\begin{align}	
 S=-{ \rm Tr} (\rho _{A} { \rm ln}\rho _{A}),
  \label{eq8}
    	\end{align}
     where $\rho_{A}$ = ${ \rm Tr}_{B} \rho $ = $ { \rm Tr}_{B}| \psi\rangle\langle\psi| $ is the reduced density matrix of one half of the chain (part \(A \), with length \(L/2 \)) after tracing out the remaining sites in part \(B \). It is a well-known physical quantity in thermodynamics that provides crucial information about the entanglement structure and quantum coherence of the system\cite{theveniaut2020transition,khemani2017critical}.
      Since the total spin projection on the \(z\) axis is conserved, we consider the largest nontrivial sector of the Hamiltonian with $\sum S_{i}^{z}=0$ (the subsequent text follows the same selection of subspace). We consider about $N=30$ eigenenergies in the window of $\varepsilon\approx0.5$. $\varepsilon=(E-E_{\operatorname*{min}})/(E_{\operatorname*{max}}-E_{\operatorname*{min}})$, where $E_{\mathrm{min}}$ ($E_{\mathrm{max}}$ ) is the lowest (highest) eigenvalue for a given disorder realization\cite{baygan2015many,orell2019probing,khemani2017critical}. We average  over the chosen eigenstates and disorder realisations to get the mean entanglement entropy $\langle S\rangle$, and plot $\langle S\rangle$ as a function of the standard deviation $ \sigma $ and the expected value $ \mu $ of the Gaussian disorder (Fig.\ref{fig3}\textcolor{blue}{(a)}).
      
      The results show that as $ \sigma $ increases, the mean entanglement entropy gradually decreases, indicating a reduction in the entanglement between the two parts of the system. This suggests that the Gaussian-disordered system undergoes a phase transition from the ergodic phase to the localized phase. Another notable observation is that the mean entanglement entropy does not vary with $ \mu $. So, this reveals that the breadth of the disorder distribution is the primary factor influencing the occurrence of MBL transition, while the central value does not affect it (we also demonstrate this issue from another perspective in Appendix.\ref{A}). The broader the disorder distribution (larger $ \sigma $), the more it promotes the transition from the ergodic phase to the localized phase.
      
      Therefore, we can also infer that the effects produced for the localized external fields \( h_i \) are randomly distributed in the intervals $\left[-h,h\right]$ (conventional disorder) and $\left[-h+a,h+a\right]$ are same. In this case, MBL transition occurs only when the disorder breadth \( 2h \) increases, regardless of the central value \( a \). In Appendix.\ref{B}, we have validated our inference.

      Since the expected value has no effect on the phase transition of the many-body system with Gaussian disorder, we set $ \mu=0 $, and use the standard deviation $ \sigma $ to denote Gaussian disorder in the subsequent study.
      
      The next physical quantity to be calculated is the mean gap ratio $\langle r\rangle$, which has long been a focal point in the study of MBL transition\cite{oganesyan2007localization,atas2013distribution,baygan2015many,luitz2015many,wiater2018impact}. The value of this physical quantity is $\langle r\rangle \approx 0.53 $ in the ergodic phase, where it follows Wigner-Dyson statistics showing level repulsion, and $ \langle r\rangle \approx 0.39 $ in the localized phase, following Poisson statistics that indicate a random distribution of energy levels\cite{thomson2023localization}. Furthermore, numerous studies have demonstrated that the mean gap ratio exhibits a generic D-shaped phase boundary along the spectrum, known as the mobility edge. This feature has been observed in MBL transitions as a second-order phase transition\cite{luitz2015many,kjall2014many,mondragon2015many,xu2020dynamical,wei2019investigating,wei2019investigating,devakul2015early,geissler2020mobility,de2016absence,wei2020characterization,yousefjani2023mobility}. We will demonstrate that systems with Gaussian disorder also exhibit the mobility edge.

      We consider the level spacings $\textstyle\delta_{\alpha}^{n}=\left|E_{\alpha}^{n}-E_{\alpha}^{n-1}\right|$,where ${E}_{\alpha}^{n}$ is the many-body eigenenergy of eigenstate \(n\) in sample $\alpha$. Then we obtain the ratio of adjacent gaps as:
      
	\begin{align}	
 r_{\alpha}^{n}=\operatorname*{min}\left\{\delta_{\alpha}^{n},\,\delta_{\alpha}^{n+1}\right\}/\operatorname{max}\left\{\delta_{\alpha}^{n},\,\delta_{\alpha}^{n+1}\right\},
  \label{eq9}
    	\end{align}
     and average this ratio over the chosen eigenstates and disorder realisations to get the mean gap ratio $\langle r\rangle$. We plot the mean gap ratio as a function of the Gaussian disorder $ \sigma $ and the energy density $ \varepsilon$. We consider 30 states within the energy density $\varepsilon \in\ [0.2,\ \ 0.8] $ and 10 states within the energy density $\varepsilon \in\ [0.1, 0.2) \cup (0.8, 0.9] $.
     
     From the phase diagram Fig.\ref{fig3}\textcolor{blue}{(b)}, we can first observe that Gaussian disorder also exhibits the mobility edge. This provides further confirmation of the validity of Gaussian disorder from another perspective. Additionally, as Gaussian disorder $ \sigma $ increases, the mean gap ratio decreases from 0.535 to 0.401, further illustrating the transition from the ergodic phase to the localized phase.

	\begin{figure}[t]	
  \centering
		\includegraphics[width=1\columnwidth]{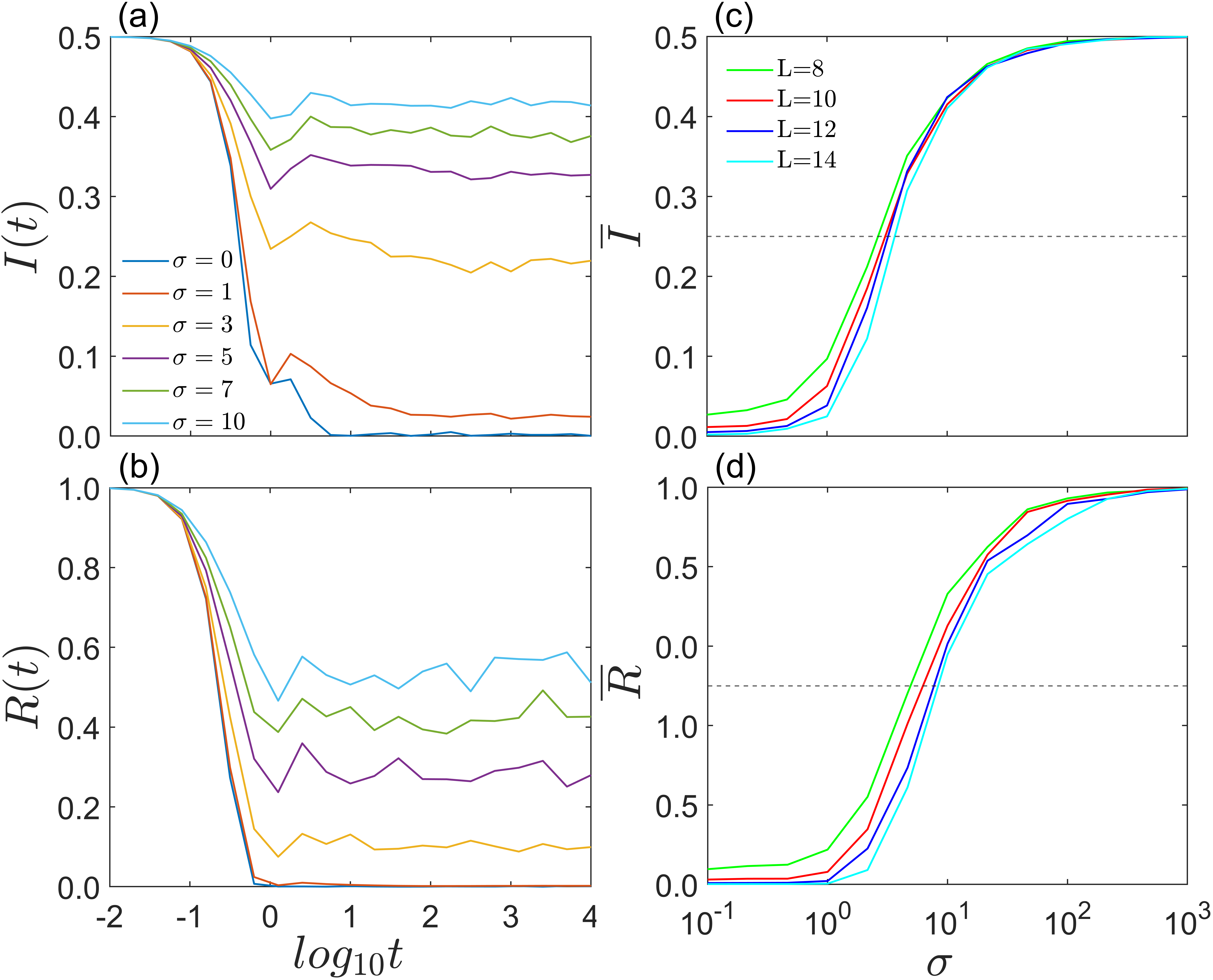}
		\caption{Time evolution of (a) the staggered magnetization \(I(t)\) and (b) the return probability \(R(t)\) starting from the Néel state for $L = 14$ with different Gaussian disorder $ \sigma $. For different chain lengths \(L\), (c) the steady-state staggered magnetization \(\overline{I}\) and (d) the steady-state return probability \(\overline{R}\) as a function of the Gaussian disorder $ \sigma $. Each point is averaged over 800 disorder realizations.}
		\label{fig4}
	\end{figure}

 \begin{figure*}[ht]
    	\centering
    	\includegraphics[width=2\columnwidth]{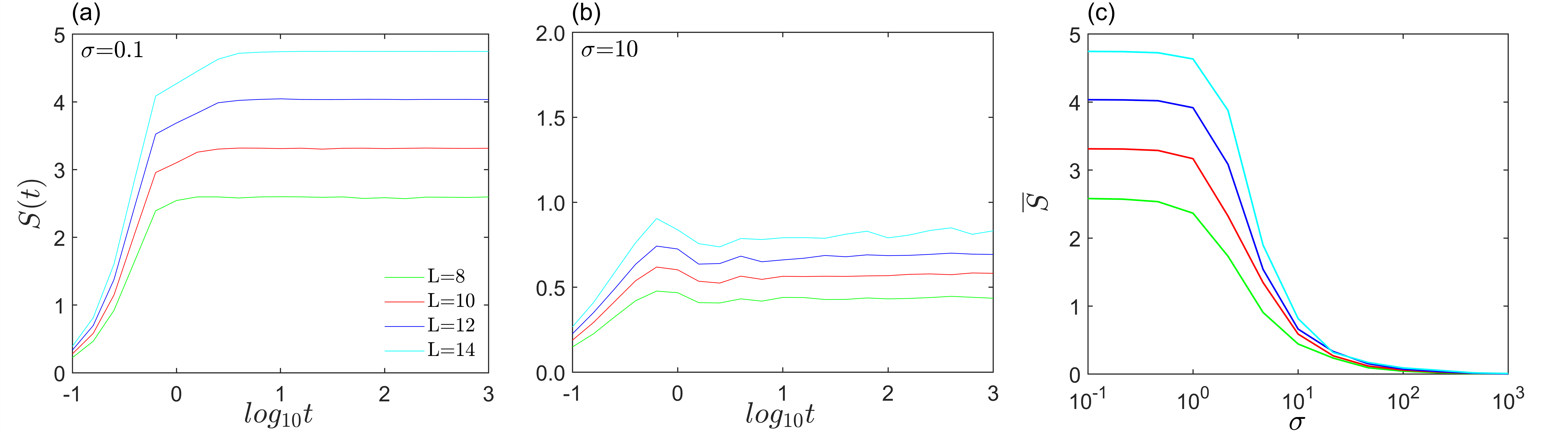}
    	\caption{ The dynamical growth of entanglement entropy \(S(t)\) of the one-dimensional spin-1/2 chain starting from the Néel state for different chain lengths \(L\) in (a) the ergodic phase and (b) the localized phase. (c) the steady-state entanglement entropy \(\overline{S}\) as a function of the Gaussian disorder $ \sigma $ for different chain lengths \(L\). Each point is averaged 800 disorder realisations.}
    	\label{fig5}
    \end{figure*}

     Next, we illustrate the dynamical characteristics of this Gaussian disordered system. We investigate the staggered magnetization following a quench from a Néel state of the form $|\uparrow\downarrow\uparrow\downarrow\uparrow\ \cdot\cdot\cdot\rangle$. The staggered magnetization defined as:
     	\begin{align}	
I(t)={\frac{1}{L}}\sum_{i=1}^{L}(-1)^{i}\left\langle\psi(t)\right|S_{i}^{z}\left|\psi(t)\right\rangle.
  \label{eq10}
    	\end{align}
      Such that \(I(0)=0.5\) and it decays in time. This observable gives us an idea of how much “memory" of its initial state the system has\cite{thomson2023localization}. If \(I(t)\) remains close to its initial value of 0.5, the system completely retains the memory of the initial state, whereas \(I(t)=0\) indicates complete loss of memory about the initial state. As a probe of localization, the staggered magnetisation has the advantage of being experimentally accessible\cite{smith2016many}, and is equivalent to the particle imbalance probed in cold-atom setups\cite{schreiber2015observation}.

      We plot the \(I(t)\) for different Gaussian disorder $ \sigma $. From the Fig.\ref{fig4}\textcolor{blue}{(a)}, it can be seen that initially \(I(0)=0.5\). Subsequently, for different Gaussian disorder $ \sigma $ values, \(I(t)\) undergoes varying degrees of decay until it reaches stability. A larger $ \sigma $ results in less decay, indicating less loss of “memory" about the initial state, which suggests a tendency towards the localized phase. Conversely, a smaller $ \sigma $ leads to more decay, indicating a tendency towards the ergodic phase.
      
      For the dynamical evolution, the final steady-state values after a quench of initial state are also worth noting. According to Fig.\ref{fig4}\textcolor{blue}{(a)}, the staggered magnetization has reached stability after a quench \((t=10^3)\) from the Néel state. So we obtain the steady-state staggered magnetization \(\overline{I}\) by averaging the time evolution over $t\in[10^{3},10^{5}] $. The Fig.\ref{fig4}\textcolor{blue}{(c)} presents \(\overline{I}\) as a function of Gaussian disorder $ \sigma $ for different chain lengths. The results show that as Gaussian disorder $ \sigma $ increases, \(\overline{I}\) increases, and when Gaussian disorder $ \sigma $ is sufficiently large, \(\overline{I}\) returns to the initial value of 0.5. This indicates that with increasing Gaussian disorder $ \sigma $, the retention of initial state information improves, and if Gaussian disorder $ \sigma $ is large enough, the initial information of the system can be well protected.

      Additionally, we calculate the return probability $\left|\langle\psi(t)|\psi(0)\rangle\right|^{2}$, which also measures how closely the quantum state retains its similarity to the initial state after time evolution\cite{wu2016understanding}. When $\left|\langle\psi(t)|\psi(0)\rangle\right|^{2}=1$ , it indicates that the quantum state remains unchanged after time evolution, meaning the system stays in the same quantum state. If $\left|\langle\psi(t)|\psi(0)\rangle\right|^{2}<1$ , it signifies that there is some difference between the quantum state after time evolution and the initial state.
      
      We plot the return probability (denoted as \(R(t)\)) for different Gaussian disorder $ \sigma $ values (Fig.\ref{fig4}\textcolor{blue}{(b)}), starting from the Néel state. When Gaussian disorder $ \sigma $ is small, \(R(t)\) quickly decays and eventually stabilizes at a small value, as $ \sigma $ increases, the stable value of \(R(t)\) after evolution gradually increases, which is consistent consistent with the behavior of the dynamics of staggered magnetization. We calculate the steady-state return probability \(\overline{R}\) in the same way as we get \(\overline{I}\). Fig.\ref{fig4}\textcolor{blue}{(d)} shows the variation of \(\overline{R}\) with Gaussian disorder $ \sigma $ for different chain lengths, as $ \sigma $ increases, \(\overline{R}\) increases, and one also can see that the initial information of the system is well protected when Gaussian disorder $ \sigma $ is large enough. The trend of the results is similar to the steady-state staggered magnetization.
      
      Moreover, according to the Fig.\ref{fig4}\textcolor{blue}{(c)} and \textcolor{blue}{(d)}, we find that the two measures of initial state preservation all have a sigmoid-shaped dependence on Gaussian disorder $ \sigma $. By thresholding at the halfway point of the sigmoid curve (indicated by the dotted line), we can roughly obtain the critical Gaussian disorder $ \sigma_{c} $ for each measure at finite chain length, and the critical point shows an increasing trend with the length of the chain. This means that as the chain length increases, the system requires stronger disorder to drive the occurrence of MBL. What is more, it can be seen that the critical point for ($L=14 $ ) is approximately consistent with the results in Fig.\ref{fig3}, 
  
    Finally, we compute the time evolution of the von Neumann entanglement entropy \(S(t)\) across a cut in the center of the chain:
      \begin{align}	
      S(t)=-\mathrm{Tr}[\,\rho_{A}(t)\mathrm{ln}(\rho_{A}(t))],
  \label{eq11}
    	\end{align}
       where $\rho_{A}(t)=\mathrm{Tr}_{B}[|\psi(t)\rangle\langle\psi(t)|]$. The entanglement entropy has previously been observed to logarithmically increase over time in the MBL phase, contrasting sharply with the much faster growth expected in the ergodic phase\cite{bardarson2012unbounded,orell2019probing,igloi2012entanglement,sierant2023slow}. It serves as a crucial indicator of the slow dynamical characteristic of MBL transition. 

      Based on the calculations of several physical quantities above, it is clear that the system is in the ergodic phase at Gaussian disorder $ \sigma=0.1 $ and in the localized phase at Gaussian disorder $ \sigma=10 $. We calculate the growth of the entanglement entropy following a quench from the initial Néel state for different chain lengths, respectively, when $ \sigma=0.1 $ and $ \sigma=10 $. From the Fig.\ref{fig5}\textcolor{blue}{(b)}, it can be observed that the middle-cut entanglement entropy of the system grows logarithmically with time when $ \sigma=10 $. The growth is faster when $ \sigma =0.1$ (Fig.\ref{fig5}\textcolor{blue}{(a)}). The above results are consistent with the theory.

      Then, we compute the variation of the steady-state entanglement entropy \(\overline{S}\) with Gaussian disorder $ \sigma $ for different chain lengths, following the procedure used for calculating \(\overline{I}\). The results show that as Gaussian disorder $ \sigma $ increases, \(\overline{S}\) decreases gradually, reducing to 0 when Gaussian disorder $ \sigma $ is sufficiently large (Fig.\ref{fig5}\textcolor{blue}{(c)}). Theses also indicate that as Gaussian disorder $ \sigma $ increases, the system transitions from the ergodic phase to the localized phase, and if Gaussian disorder $ \sigma $ is large enough, the system reaches deep localization, the initial information of the system is well protected.

      The above results all indicate that the Gaussian disorder can induce the MBL transition. However, what plays a decisive role in this process is the standard deviation $ \sigma $, not the expected value $ \mu $. By analyzing the properties of eigenstates and dynamical characteristics of the spin-1/2 chains with Gaussian disorder, it is shown that as the standard deviation $ \sigma $ increases, the system gradually transitions from the ergodic phase to the localized phase. Studying this type of disorder helps us gain deeper insights into the MBL transition induced by disordered external fields, where the crucial factor is the breadth of the disorder rather than its central value.

   \section{\label{sec4}The mutual influence between the system and the environment}
   
	\begin{figure}[t]	
  \centering
		\includegraphics[width=0.8\columnwidth]{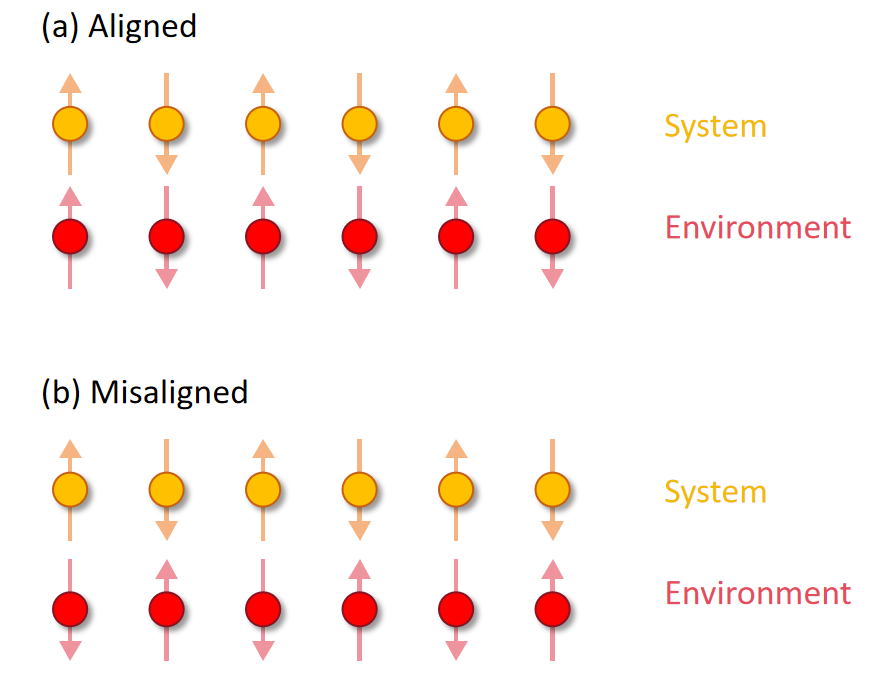}
		\caption{For the sys-env models, there are two different initial states: (a) aligned state, where the spins of the nearest neighboring sites in the upper and lower chain are aligned in the same direction. (b) misaligned state, where the spins of the nearest neighboring sites in the upper and lower chain are aligned in opposite directions.}
		\label{fig6}
	\end{figure}

     In this section, we apply the Gaussian disorder to the sys-env model, considering the mutual influence between the system and environment when the lengths of the them are comparable. 
     
     Our first sys-env model in ladder configuration (introduced in Sec.\ref{ladder mod}) is similar to the model from Chiew's previous research\cite{chiew2023stability}, but with differences. Specifically, we apply different types of disorder to each part: conventional disorder in the system and Gaussian disorder in the environment. Additionally, we have considered next-nearest-neighbor interactions within both chains. We focus on analyzing the mutual influence between the system and the environment. Relevant research is discussed in Sec.\ref{subsec1}.

     Additionally, we advance the sys-env model in ladder configuration to staggerd configuration (introduced in Sec.\ref{stragged mod}), which simulates more complex coupling situations between the system and the external environment. The mutual influence between them in this model are presented in Sec.\ref{subsec2}.

    The indicator we use is the steady-state staggered magnetization \(\overline{I}\) (Appendix.\ref{C} provides another indicator). The staggered magnetizations of the system and environment are:
           \begin{align}	
I_{sys}\left(t\right)&=\frac1{l}\sum_{i=1}^{l}\left(-1\right)^{i}\left<\psi(t)\right|S_{i}^{z}\left|\psi(t)\right>,\\I_{env}\left(t\right)&=\frac1{l}\sum_{i=l+1}^{2l}\left(-1\right)^{i}\left<\psi(t)\right|S_{i}^{z}\left|\psi(t)\right>.
  \label{eq12}
    	\end{align}

     $ \left|\psi(t)\right>=e^{-i H_{tot} t}\left|\psi(0)\right>$ is the state obtained after the evolution of the initial state $|\psi(0)\rangle$ for time \(t\). The steady-state staggered magnetization of the system \(\overline{I}_{sys}\) and the environment \(\overline{I}_{env}\) are used respectively to quantify the extent of preservation of their initial state information. \(\overline{I}_{sys}=0.5\) (\(\overline{I}_{env}=0.5\)) indicates that the system (environment) retains its initial state information. \(\overline{I}_{sys}=0\) (\(\overline{I}_{env}=0\)) indicates that initial state information of the system (environment) is completely lost. We consider two different initial states: the aligned state (Fig.\ref{fig6}\textcolor{blue}{(a)}) and the misaligned state (Fig.\ref{fig6}\textcolor{blue}{(b)}). The interaction terms \(H_{int}\) lead to different energy densities for the two initial states.

\subsection{\label{subsec1}Ladder configuration} 
\begin{figure}[t]
    	\centering
    	\includegraphics[width=1\columnwidth]{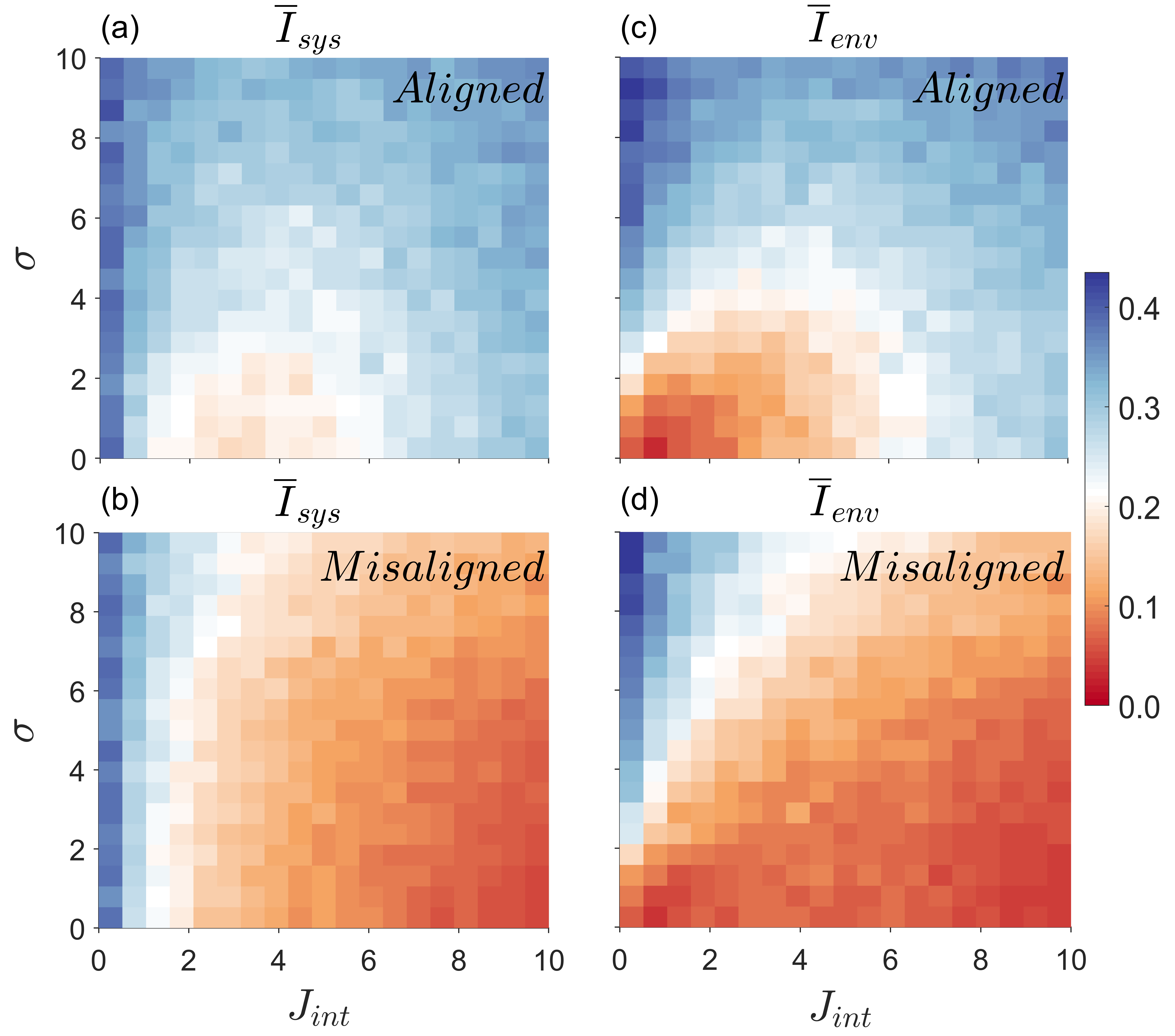}
    	\caption{ The conventional disorder of the system $h=10J$. Steady-state staggered magnetisation for the system \(\overline{I}_{sys}\) (left) and the environment \(\overline{I}_{env}\) (right) for the ladder configuration. The upper two plots are for the initial aligned state. The bottom two plots are for the initial misaligned state. Each point is averaged 100 disorder realisations.}
    	\label{fig7}
    \end{figure}

    \begin{figure}[t]
    	\centering
    	\includegraphics[width=1\columnwidth]{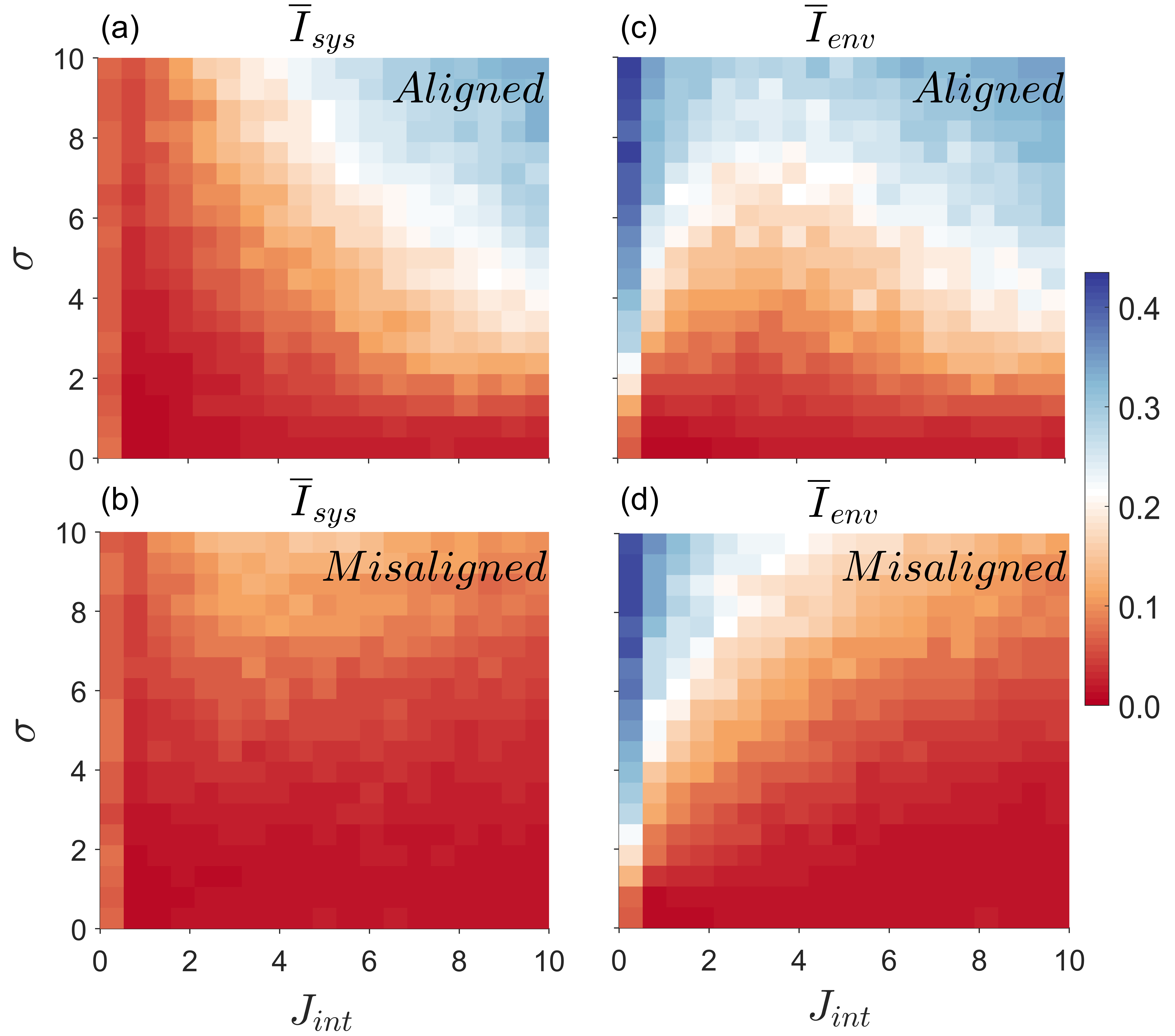}
    	\caption{ The conventional disorder of the system $h=0.5J$. Steady-state staggered magnetisation for the system \(\overline{I}_{sys}\) (left) and the environment \(\overline{I}_{env}\) (right) for the ladder configuration. The upper two plots are for the initial aligned state. The bottom two plots are for the initial misaligned state. Each point is averaged 100 disorder realisations.}
    	\label{fig8}
    \end{figure}
     We separately set the conventional disorder of the system to be $ h=10J$ and $ h=0.5J$, so that the system initially lies well in the localized and the ergodic phase, while allowing Gaussian disorder $ \sigma$ of the environment and the coupling strength \( J_{int} \) between the system and the environment to vary from 0 to 10. 

     Fig.\ref{fig7} and Fig.\ref{fig8} respectively show the results when the system is initially in the localized and the ergodic phase. The left (right) plots represent the steady-state staggered magnetisation of the system (environment) \(\overline{I}_{sys}\) (\(\overline{I}_{env}\)), with the upper (bottom) plots corresponding to the initial state being in the aligned (misaligned) state.

    When $J_{i n t}\ll J$ (corresponding to $J_{i n t}\approx 0$ in Fig.\ref{fig7} and Fig.\ref{fig8}), the interaction strength between two chains can be considered negligible compared to the interaction strength within each chain. From Fig.\ref{fig7}\textcolor{blue}{(a) (b)} and Fig.\ref{fig8}\textcolor{blue}{(a) (b)}, the system's retention of its initial information depends on its own conventional disorder \( h \) and is not affected by the environmental Gaussian disorder $\sigma $. As \( h \) decreases sequentially from 10\( J \) to 0.5\( J \), \(\overline{I}_{sys}\) also decreases from 0.41 to 0.07. From Fig.\ref{fig7}\textcolor{blue}{(c) (d)} and Fig.\ref{fig8}\textcolor{blue}{(c) (d)}, as the environmental Gaussian disorder $\sigma $ increases, the environment's retention of initial state information continuously improves and is not affected by the system's Gaussian disorder \( h \). This can be explained as the interaction between the system and the environment being very weak, so they behave like two independent chains. Their dynamic behavior is determined by their initial characteristics. The degree to which the system and environment retain initial state information depends on their own disorder strength. They do not affect each other.

    In the case of $J_{i n t}\approx J$ (corresponding to $J_{i n t}\approx 1$ in Fig.\ref{fig7} and Fig.\ref{fig8}), the system and environment are structurally connected. From Fig.\ref{fig7}\textcolor{blue}{(a) (b)}, we can see that when the localized system is coupled with the environment, it can still relatively retain the initial state information. Additionally, as the environmental Gaussian disorder $\sigma $ increases, the extent of information retention of the initial state slightly improves. Due to the interchangeability between the system and the environment, we can observe similar results in the environment. For the environment, Fig.\ref{fig7}\textcolor{blue}{(c) (d)} and Fig.\ref{fig8}\textcolor{blue}{(c) (d)} show that the degree of information retention of the initial state in the environment increases with the increase in its own disorder. On the other hand, by comparing Fig.\ref{fig7}\textcolor{blue}{(c) (d)} and Fig.\ref{fig8}\textcolor{blue}{(c) (d)} , we can see that the environment is also affected by the disorder of the system. As the system's conventional disorder decreases from 10\( J \) to 0.5\( J \), the degree of information retention in the environment decreases, though the change is relatively minor. Overall, the connection established between the them cannot be ignored and they have started to influence each other. However, the intrinsic disorder still plays a dominant role.

    In the case of $J_{i n t}\gg J$ (corresponding to larger \( J_{int} \) values in Fig.\ref{fig7} and Fig.\ref{fig8}), compare the upper and bottom plots in Fig.\ref{fig7} or Fig.\ref{fig8}, we can observe that the effects caused by different initial states gradually become apparent. When \( J_{int} \) exceeds 6, for the same initial state, the degree to which the system and environment retain initial state information shows a strong correlation ${\overline{I}}_{sys}\approx{\overline{I}}_{env}$, and this degree depends on the initial state. Chiew's analysis for this difference caused by the initial state is also applicable here\cite{chiew2023stability}. Although we have additionally considered the next-nearest-neighbor coupling within each chain , when \( J_{int} \) is strong, the spin particles of a rung can still be treated as a dimer. Different initial states have different energy densities, leading to different outcomes. We will discuss the impact of the initial state in more detail in Sec.\ref{sec5}. In this section, we primarily focus on the mutual influence between them. From the comparison of the trends of \(\overline{I}_{sys}\) at the right edges of Fig.\ref{fig7}\textcolor{blue}{(a) (b)} and Fig.\ref{fig8}\textcolor{blue}{(a) (b)} with change in environment disorder, it can be observed that the system in the localized phase is less affected by environmental disorder than the system in the ergodic phase. Similarly, the effect is the same for the environment. The environment in the localized phase (upper right corners of the four \(\overline{I}_{env}\) in Fig.\ref{fig7} and Fig.\ref{fig8}) shows a less response to the system's disorder changing from 10\( J \) to 0.5\( J \) compared to the environment in the ergodic phase (lower right corners of the four \(\overline{I}_{env}\) in Fig.\ref{fig7} and Fig.\ref{fig8}). Therefore, in the scenario of strong \( J_{int} \), both the system and the environment are influenced by each other (particularly evident when they are in the ergodic phase).
    
    The trends in our results for Fig.\ref{fig7} are generally consistent with those in previous studies\cite{chiew2023stability}. Although our model differs, the difference in numerical ranges is primarily due to different methods for calculating the staggered magnetization. Additionally, for the same values, the Gaussian disorder we introduced corresponds to a broader disorder breadth compared to the conventional disorder (this will be explained in the Sec.\ref{subsec3}). Thus, our results show a “longitudinal compression" effect compared to theirs.

\subsection{\label{subsec2}Staggered configuration} 
\begin{figure}[t]
    	\centering
    	\includegraphics[width=1\columnwidth]{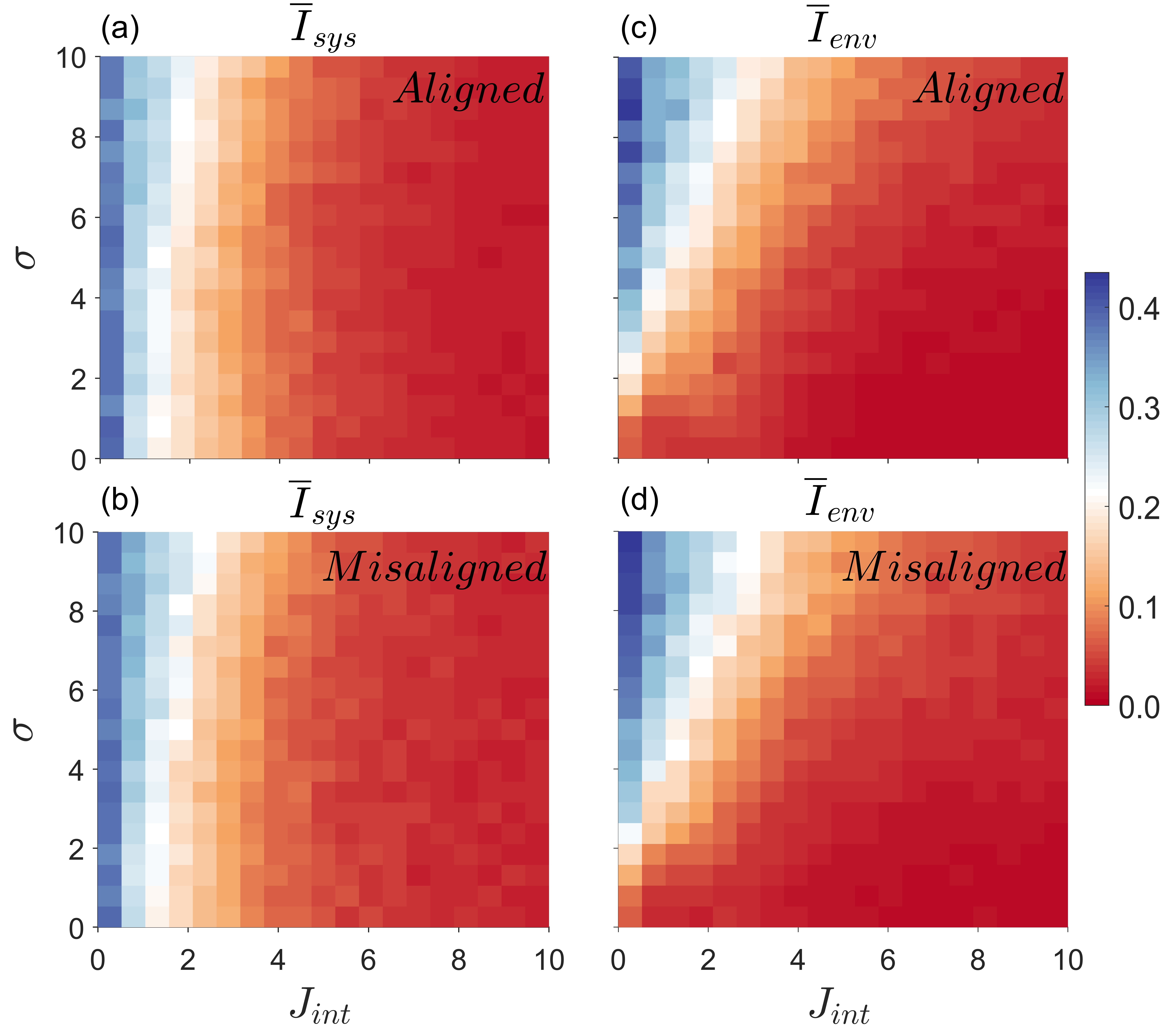}
    	\caption{ The conventional disorder of the system $h=10J$. Steady-state staggered magnetisation for the system \(\overline{I}_{sys}\) (left) and the environment \(\overline{I}_{env}\) (right) for the staggered configuration. The upper two plots are for the initial aligned state. The bottom two plots are for the initial misaligned state. Each point is averaged 100 disorder realisations.}
    	\label{fig9}
    \end{figure}
    \begin{figure}[t]
    	\centering
    	\includegraphics[width=1\columnwidth]{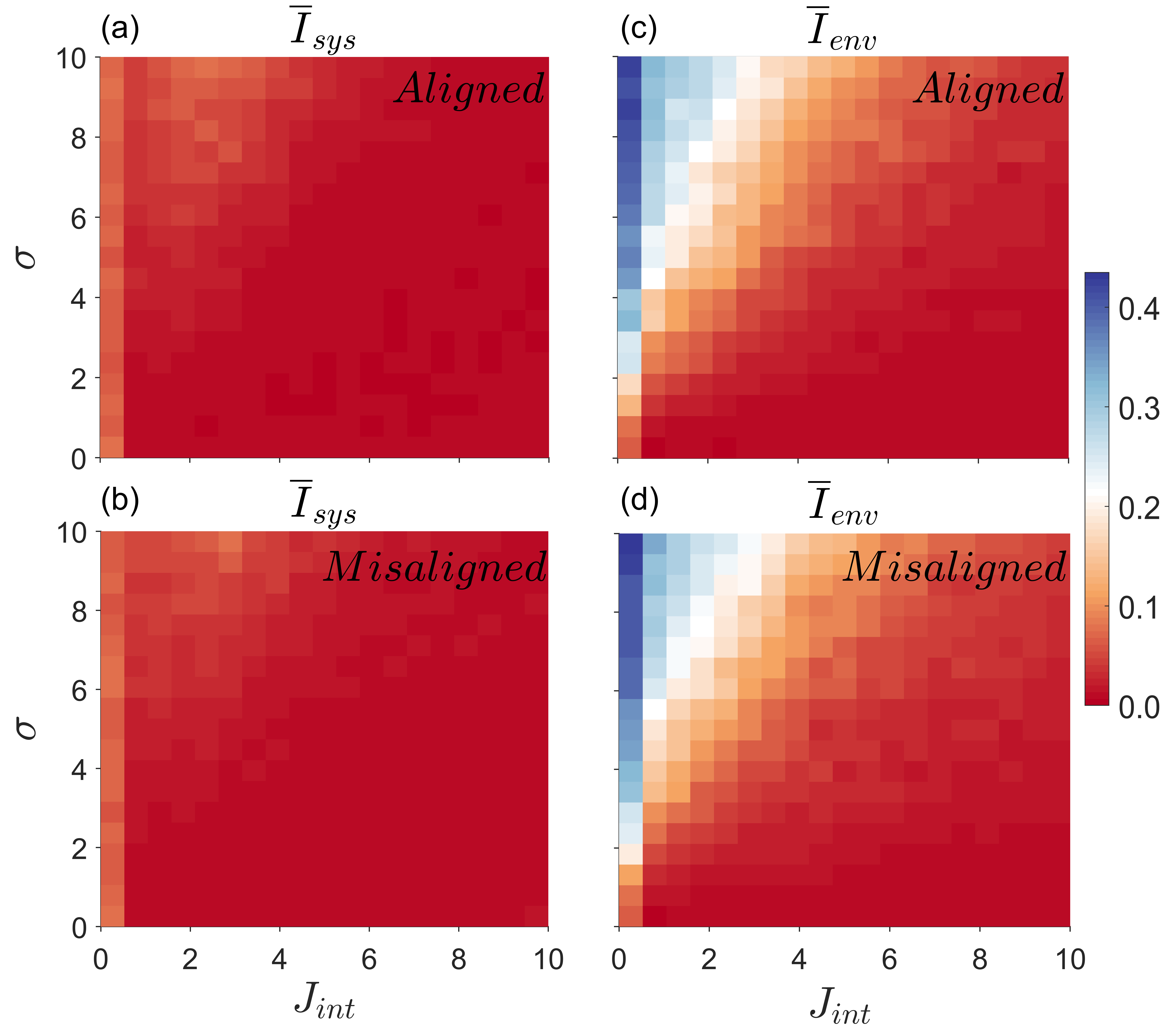}
    	\caption{ The conventional disorder of the system $h=0.5J$. Steady-state staggered magnetisation for the system \(\overline{I}_{sys}\) (left) and the environment \(\overline{I}_{env}\) (right) for the staggered configuration. The upper two plots are for the initial aligned state. The bottom two plots are for the initial misaligned state. Each point is averaged 100 disorder realisations.}
    	\label{fig10}
    \end{figure}
    
    For the sys-env model in staggered configuration, we conduct the same analysis.  The results of the initial localized system ($h=10J$) and the initial ergodic system ($h=0.5J$) are presented in Fig.\ref{fig9} and Fig.\ref{fig10}, respectively. The left (right) plots represent the steady-state staggered magnetisation of the system (environment) \(\overline{I}_{sys}\) (\(\overline{I}_{env}\)). The upper (bottom) plots corresponding to the initial state being in the aligned (misaligned) state.

   Initially, at $J_{i n t}\ll J$ (corresponding to $J_{i n t}\approx 0$ in Fig.\ref{fig9} and Fig.\ref{fig10}), the results are similar to those in the ladder configuration. From Fig.\ref{fig9}\textcolor{blue}{(a) (b)} and Fig.\ref{fig10}\textcolor{blue}{(a) (b)}, it can be observed that the system's retention of its initial information depends on its own disorder \(h\) and is not affected by the Gaussian disorder strength of the environment. Correspondingly, from Fig.\ref{fig9}\textcolor{blue}{(c) (d)} and Fig.\ref{fig10}\textcolor{blue}{(c) (d)}, it can be observed that as environmental Gaussian disorder $\sigma$ increases, the environment enhances the retention of initial state information, unaffected by the disorder \(h \) of the system. This is consistent with the results in the ladder configuration, and the reasons are also similar to those in the ladder configuration.

   When $J_{i n t}\approx J$ (corresponding to $J_{i n t}\approx 1$ in Fig.\ref{fig9} and Fig.\ref{fig10}), the system and environment have established a structural connection. From Fig.\ref{fig9}\textcolor{blue}{(a) (b)} (the upper strong $\sigma$ section in Fig.\ref{fig9}\textcolor{blue}{(c) (d)} and Fig.\ref{fig10}\textcolor{blue}{(c) (d)}), it can be seen that the localized system (environment) still retains the initial information relatively well, though it is slightly affected by the disorder $\sigma$ (\(h\)) from the other chain. From Fig.\ref{fig10}\textcolor{blue}{(a) (b)} (the bottom weak $\sigma$ section in Fig.\ref{fig9}\textcolor{blue}{(c) (d)} and Fig.\ref{fig10}\textcolor{blue}{(c) (d)}, it can be seen that the ergodic system (environment) has completely lost the information of the initial state. Moreover, comparing with the results from Sec.\ref{subsec1}, the impact of disorder from the environment (system) on the system (environment) is weaker compared to the ladder configuration. This is because, in the staggered configuration, even with the same \( J_{int} \) value, the greater number of couplings between the two chains results in a higher coupling strength, making the system and the environment insensitive to the change of both their own disorder and the other disorder than before. 

   However, what is significantly different is in the case of $J_{i n t}\gg J$ (corresponding to larger \( J_{int} \) values in Fig.\ref{fig9} and Fig.\ref{fig10}). For the staggered configuration, the initial state does not affect their eventual localization properties. Regardless of whether it is the aligned state or the misaligned state, when \( J_{int} \) exceeds approximately 6, ${\overline{I}}_{sys}\approx{\overline{I}}_{env}\approx0$. This indicates that within the range of disorder strengths we have studied, both the system and the environment have essentially lost all information of the initial state. Therefore, we can infer that in this configuration, the differences in energy density arising from different initial states are extremely small and can be neglected. Additionally, due to the strong and complex coupling between the two chains, they are less sensitive to changes in their own disorder as well as the disorder from the other chain. Thus, within the range of disorder strengths we studied, they essentially do not affect each other. This situation can be understood as when both the system and the environment strongly interfere with each other, both will completely lose their initial information.

\section{\label{sec5}Many-body localization of the entire system.} 
   In this section, we study the localization of the entire system composed of the system and the environment, and emphasize the necessity of careful interpretation when analyzing the system using dynamic indicators. The total Hamiltonian remains \( H_{tot} \). The total chain length is 12, with sites 1-6 denoting the system spins and sites 7-12 denoting the environment spins.

\subsection{\label{subsec3}Ladder configuration} 

  \begin{figure*}[ht]
  \centering
		\includegraphics[width=2\columnwidth]{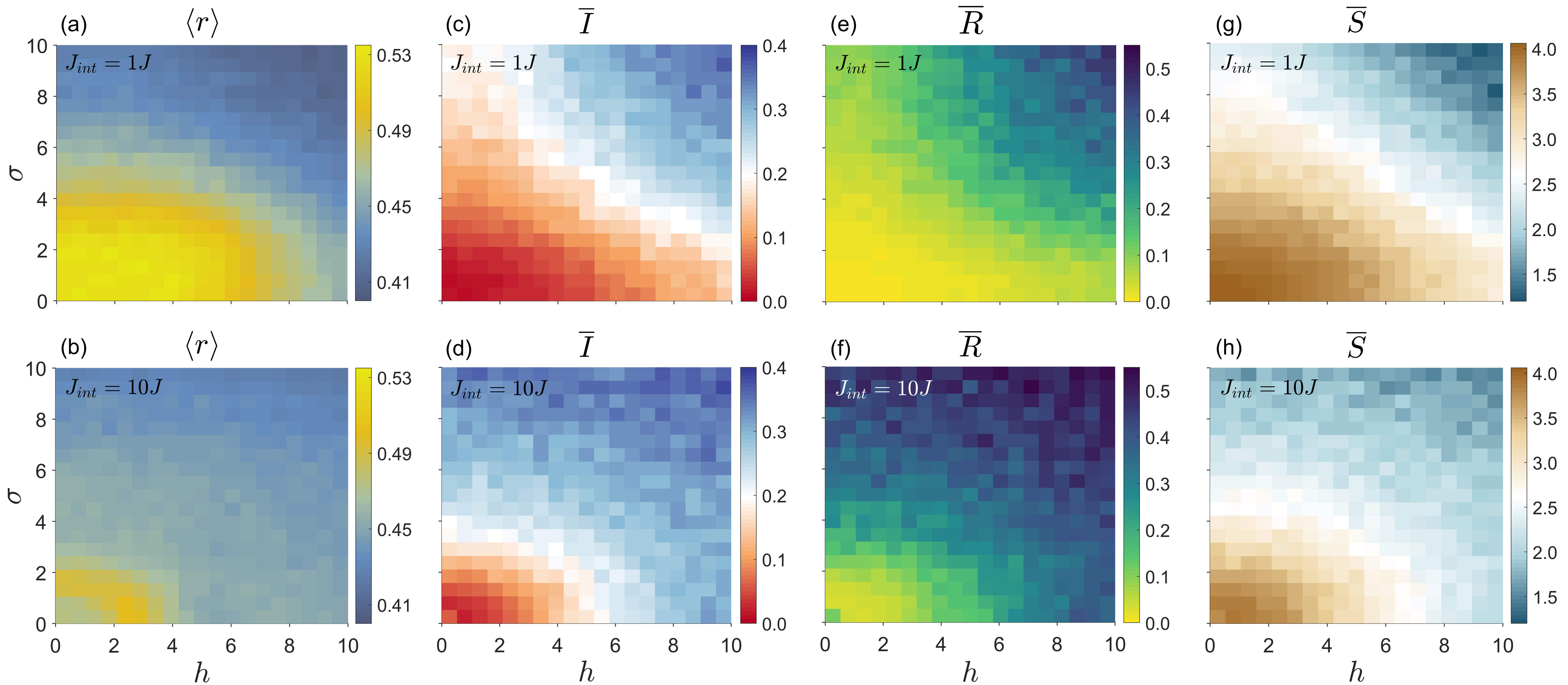}
		\caption{In the ladder configuration, the mean gap ratio $\langle r\rangle$ (first column, selecting 30 states near an energy density of $\varepsilon\approx0.5$ as samples), the steady-state staggered magnetization \(\overline{I}\) (second column), the steady-state return probability \(\overline{R}\) (third column), and the steady-state entanglement entropy \(\overline{S}\) (fourth column) of the entire system vary with Gaussian disorder $ \sigma $ and conventional disorder \( h \), for different values of \( J_{int} \). The upper four plots, $ J_{int} = 1J$. The bottom four plots, $ J_{int} = 10J$. Each point is averaged 100 disorder realisations.}
		\label{fig11}
	\end{figure*}

    Firstly, we examine how the disorder strength of each chain affect the localization of the entire system. We calculate the mean gap ratio $\langle r\rangle$, which reflects the properties of eigenstates of the many-body system, as well as three dynamical steady-state quantities: the steady-state staggered magnetization \(\overline{I}\), the steady-state return probability \(\overline{R}\), and the steady-state entanglement entropy \(\overline{S}\), with $|\uparrow\downarrow\uparrow\downarrow\uparrow\downarrow\uparrow\downarrow\uparrow\downarrow\uparrow\downarrow\rangle$ as the initial state (corresponding to the aligned state mentioned earlier). We fix $ J_{int} = 1J$ and $ J_{int} = 10J$, varying the conventional disorder \(h\) of the upper chain and Gaussian disorder $ \sigma $ of the lower chain (Fig.\ref{fig11}). From the results of $\langle r\rangle$ (Fig.\ref{fig11}\textcolor{blue}{(a) (b)}), as the conventional disorder \(h\) and Gaussian $ \sigma $ increases, the entire system gradually approaches the localized phase. The other three dynamical steady-state indicators also yield consistent results (Fig.\ref{fig11}\textcolor{blue}{(c)-(h)}). Additionally, from Fig.\ref{fig11}\textcolor{blue}{(a) (b)}, it is evident that even when disorder in one chain (either \( h \) or $ \sigma $) is small, the MBL transition can still occur if the disorder in the other chain is sufficiently large. This also indicates that the localization is jointly regulated by the disorder of both parts. Furthermore, the asymmetry in the each plot of Fig.\ref{fig11} is attributed to the different forms of disorder present in the two parts. This implies that for the same value of $ \sigma $ and \( h \), the disorder breadth of the former is greater than the latter, hence the plots in Fig.\ref{fig11} is more sensitive to changes in $ \sigma $ than \( h \).

\begin{figure}[t]	
  \centering
		\includegraphics[width=1\columnwidth]{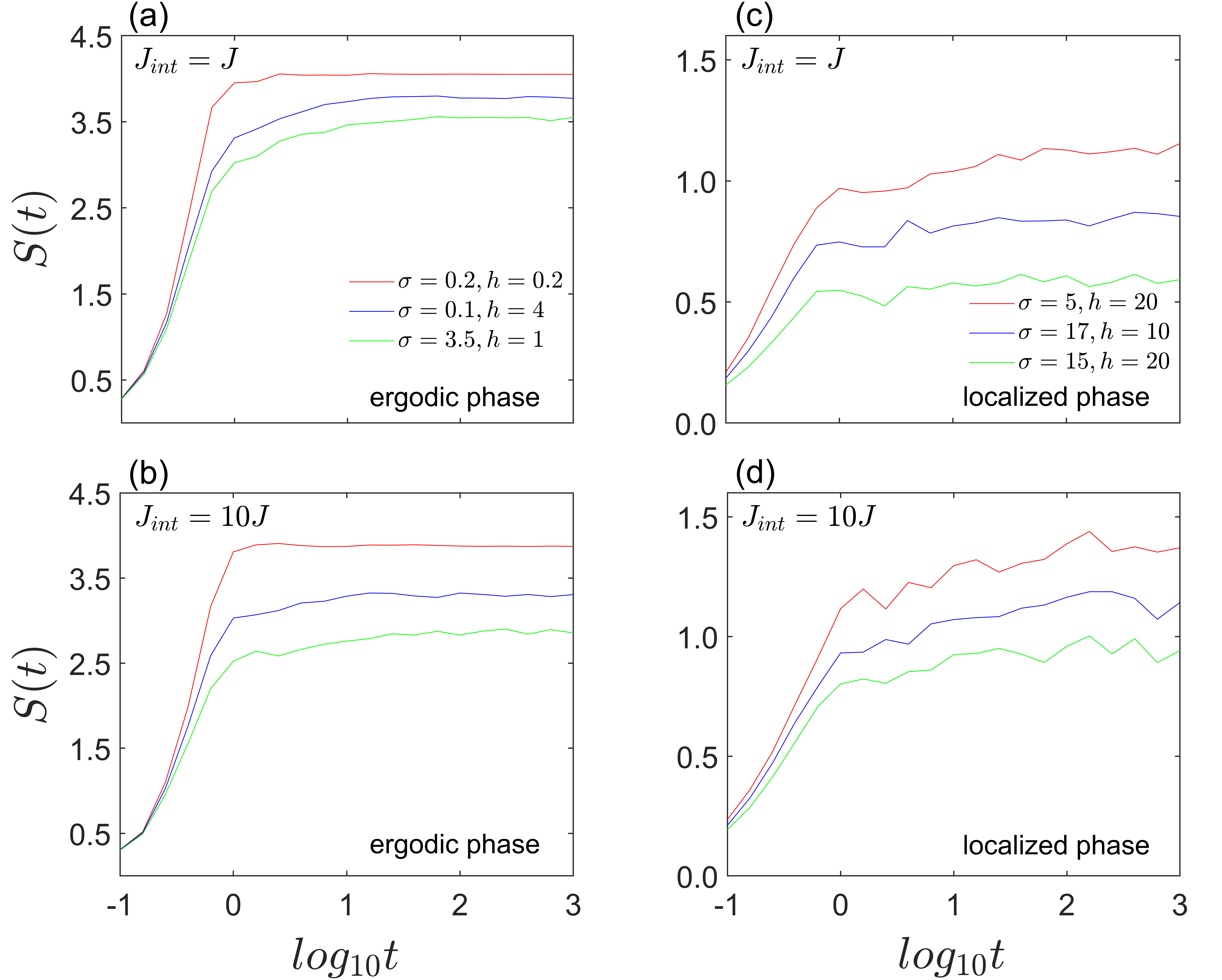}
		\caption{In the ladder configuration, the dynamical growth of entanglement entropy \(S(t)\) starting from the state $|\uparrow\downarrow\uparrow\downarrow\uparrow\downarrow\uparrow\downarrow\uparrow\downarrow\uparrow\downarrow\rangle$ for different combinations of $ \sigma $ and \( h \) with different values of \( J_{int} \). (a) $ J_{int} = 1J$, weak disorder. (b) $ J_{int} = 1J$, strong disorder. (c) $ J_{int} = 10J$, weak disorder. (d) $ J_{int} = 10J$, strong disorder. Each point is averaged 1000 disorder realisations.}
		\label{fig12}
	\end{figure}

    Then we distinguish the localized phase and the ergodic phase through the evolution of the mid-cut entanglement entropy \(S(t)\) of the entire system (Fig.\ref{fig12}). Based on the phase diagram of $\langle r\rangle$, still for $ J_{int} = 1J$ and $ J_{int} = 10J$, we selecte some combinations of different values of Gaussian disorder $ \sigma $ and conventional disorder \( h \), and plot the \(S(t)\) of the entire system in the localized phase and the ergodic phase separately. At weak disorder, the entanglement entropy grows rapidly and quickly saturates, whereas at strong disorder, it has a slow, logarithmic growth and then saturates.

 \begin{figure}[t]	
  \centering
		\includegraphics[width=1\columnwidth]{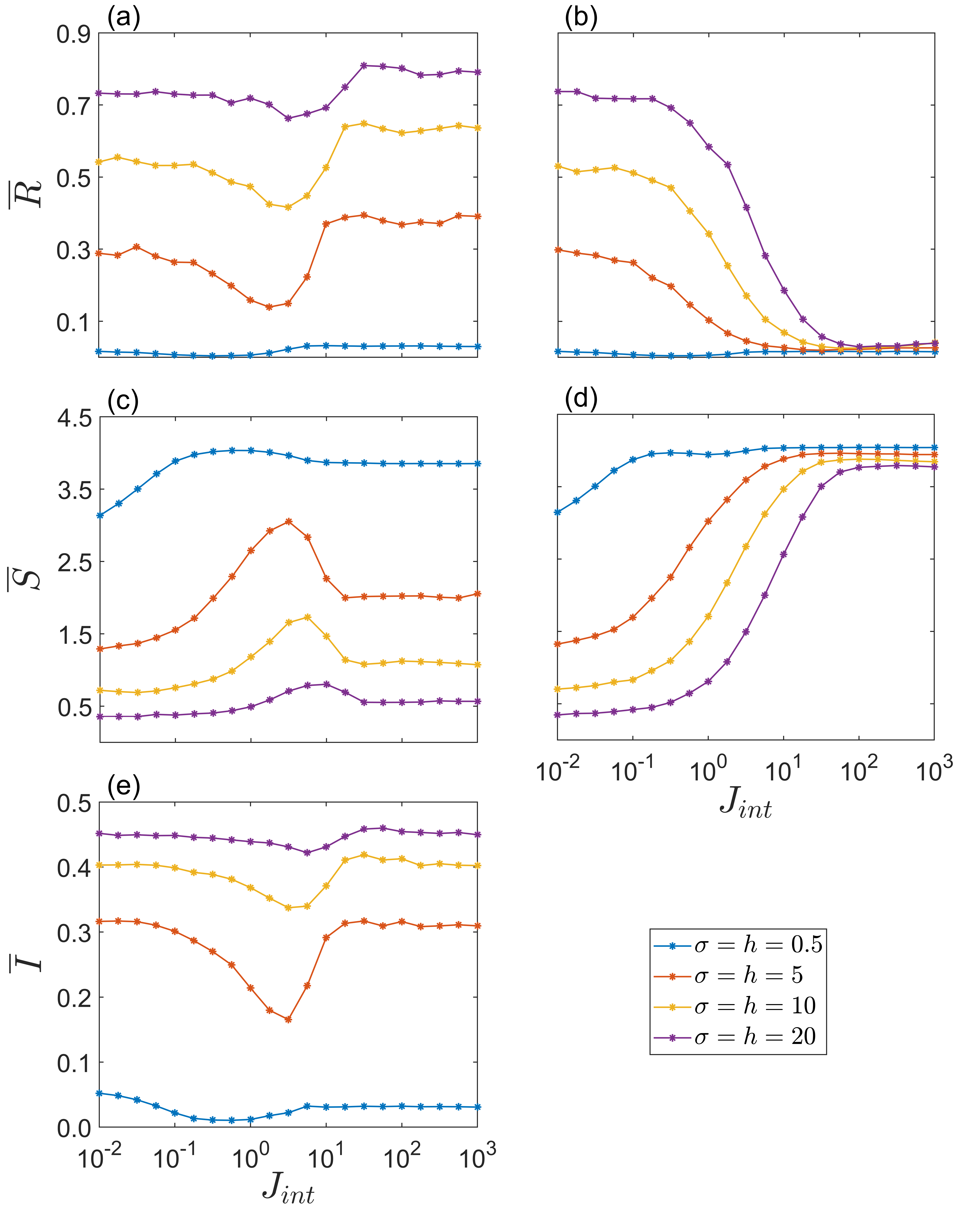}
		\caption{In the ladder configuration, (a) (b) the steady-state return probability \(\overline{R}\), (c) (d) the steady-state entanglement entropy \(\overline{S}\), and (e) the steady-state staggered magnetization \(\overline{I}\) vary with \( J_{int} \) for different values of disorder. The three plots on the left are for the initial aligned state. The two plots on the right are for the initial misaligned state. Each point is averaged 1000 disorder realisations.}
		\label{fig13}
	\end{figure}
 
   In Sec.\ref{sec4}, we analyze the effect of \( J_{int} \) on the mutual influence between the system and the environment. Therefore, here we also investigate how the dynamical characteristics of the entire system change with \( J_{int} \). And as Sec.\ref{sec4} shown, different initial states may lead to different results. Thus, we speculate whether the dynamic indicators used to analyze the characteristics of the entire system will also be affected by the initial state. We analyze this issue by selecting two different initial states: $|\uparrow\downarrow\uparrow\downarrow\uparrow\downarrow\uparrow\downarrow\uparrow\downarrow\uparrow\downarrow\rangle$ and $|\uparrow\downarrow\uparrow\downarrow\uparrow\downarrow\downarrow\uparrow\downarrow\uparrow\downarrow\uparrow\rangle$, which correspond to the previous aligned state and misaligned state, respectively. We use the dynamical steady-state indicators, which are the steady-state staggered magnetization, the steady-state return probability, and the steady-state entanglement entropy. Because the staggered magnetization depends upon the initial state being in the Néel form, we only use the initial state $|\uparrow\downarrow\uparrow\downarrow\uparrow\downarrow\uparrow\downarrow\uparrow\downarrow\uparrow\downarrow\rangle$ for steady-state staggered magnetization. We plot the steady-state return probability (Fig.\ref{fig13}\textcolor{blue}{(a) (b)}), the steady-state entanglement entropy (Fig.\ref{fig13}\textcolor{blue}{(c) (d)}), and the steady-state staggered magnetization (Fig.\ref{fig13}\textcolor{blue}{(e)}) of the entire system as functions of the coupling strength \( J_{int} \) between the two chains. $ \sigma=h=10$ and $ \sigma =h=0.5$ in Fig.\ref{fig13} correspond to the area above in Fig.\ref{fig7} ($\sigma =10$) and the area below in Fig.\ref{fig8} ($\sigma =0.5$), respectively. We find that the trends of these steady-state indicators differ for the two different initial states, particularly after \( J_{int} \) is around $10^{0}$.

  As \( J_{int} \) increases from $10^{-2}$ to $10^{0}$, all steady-state indicators suggest that the entire system shows a gradual trend towards thermalization, regardless of whether the initial state is aligned and misaligned, which is consistent with the results in Sec.\ref{subsec1}. The weakening of localization can be explained as the entire system transitions from two mutually independent single chains to a quasi-two-dimensional configuration that requires stronger disorder to localize..

  When \( J_{int} \) increases from $10^{0}$ to $10^{3}$, for the aligned initial state, all indicators show a turning point at $J_{i n t}\approx 10^{0}$, displaying a “rebound" trend and eventually stabilizing around $J_{i n t}\approx 10^{2}$, while, for the misaligned initial state, the indicators continue to show a trend towards thermalization. These results are also consistent with the previous results in Sec.\ref{subsec1}. Here, different initial states yield different results, we can explain it as follow. 
  
  For the initial aligned state, at strong \( J_{int} \) , each dimer of a rung is in a spin-1 mode, so the ladder is reduced to a chain of interacting spin-1 particles (this conclusion has been confirmed in the study referenced as \cite{chiew2023stability}). The steady-state values are qualitatively equivalent to those of a disordered spin-1 chain. Therefore, due to the transition from a two-dimensional ladder to a one-dimensional model, the indicators show a “rebound" when the disorder strength is kept constant. After \( J_{int} \) reaches $10^{2}$ ($J_{i n t}\gg J$), all indicators no longer change, indicating that the system has fully transitioned to a single chain. Additionally, all indicators consistently indicate that with increasing disorder strength, the “turning point” shifts backward, suggesting that that the localized system requires a larger \( J_{int} \) to drive the model's transition. Furthermore, the steady-state staggered magnetization (Fig.\ref{fig13}\textcolor{blue}{(e)}) calculated for Néel initial state is consistent with the results obtained for the steady-state staggered magnetization of system (Fig.\ref{fig7}\textcolor{blue}{(a)} and Fig.\ref{fig8}\textcolor{blue}{(a)}) and the steady-state staggered magnetization of environment (Fig.\ref{fig7}\textcolor{blue}{(b)} and Fig.\ref{fig8}\textcolor{blue}{(b)}) in Sec.\ref{subsec1}. For the initial aligned state, it is evident that the eventually localization properties of both the system and the environment exhibit a trend of initially decreasing and then increasing, making it reasonable for the entire system to show a similar trend. From Fig.\ref{fig13}\textcolor{blue}{(a)} and \textcolor{blue}{(b)}, it can be seen that the results of the local properties of the entire system presented by the other two quantities are also consistent.

  For the initial misaligned state, each dimer oscillates between spin-0 and spin-1 modes, so the ladder is reduced to a single spin chain consisting of particles that have both spin-0 and spin-1 degrees of freedom (this conclusion has been confirmed in the study referenced as \cite{chiew2023stability}). The result can be understood as the local oscillations driven by inhomogeneity in energy scales in the Hamiltonian consequently lead to subsequent dynamics that appear to thermalize (information loss). The two physical quantities: the steady-state return probability (Fig.\ref{fig13}\textcolor{blue}{(b)}) and the steady-state entanglement entropy (Fig.\ref{fig13}\textcolor{blue}{(d)}), we used both exhibit a similar trend towards thermalization. This is consistent with the trend observed in Sec.\ref{subsec1}, where both the system (Fig.\ref{fig7}\textcolor{blue}{(c)} and Fig.\ref{fig8}\textcolor{blue}{(c)}) and environment (Fig.\ref{fig8}\textcolor{blue}{(d)} and Fig.\ref{fig8}\textcolor{blue}{(d)}) showed a similar thermalization trend when the initial state is misaligned.

  So, for certain system structure (such as the strong-coupling ladder configuration here) with the same disorder strength, different initial states can even produce drastically different dynamical results. This is because, in such configuration, differences in initial states can lead to significant variations in energy density, resulting in different localization properties. This is consistent with the reasons that cause the MBL mobility edges\cite{luitz2015many,kjall2014many,mondragon2015many,xu2020dynamical,wei2019investigating,wei2019investigating,devakul2015early,geissler2020mobility,de2016absence,wei2020characterization,yousefjani2023mobility}, where states at different energy densities may exhibit different localization properties. Therefore, for certain systems, we cannot determine the localization properties of the system solely based on its dynamical characteristic from some initial staes. Actually, the dynamical results for the initial state only reflect the degree to which the system retains information of this initial state. This means that a localized system cannot retain information for all initial states, where our aligned and misaligned states serve as examples of retainable and non-retainable information, respectively.

  In Fig.\ref{fig11}\textcolor{blue}{(a)} and \textcolor{blue}{(b)}, we study the properties of eigenstates of the entire system, calculating the mean gap ratio $\langle r\rangle$ that accurately reflects the localization properties of the many-body system. The results show that compared to $ J_{int} = 1J$ (Fig.\ref{fig11}\textcolor{blue}{(a)}), the entire system exhibits stronger localization properties when $ J_{int} = 10J$ (Fig.\ref{fig11}\textcolor{blue}{(b)}). From Fig.\ref{fig11}\textcolor{blue}{(c)-(f)}, we can find that, for the initial aligned state, the results shown by the other three dynamical indicators are consistent with $\langle r\rangle$. Therefore, in this configuration, the dynamical results of the aligned state can accurately reflects the localization properties of the entire system, which is why we only calculate the dynamical indicators for the aligned state in Fig.\ref{fig11}\textcolor{blue}{(c)-(h)}. Additionally, as shown on the left side of Fig.\ref{fig13}, the aligned state can intuitively demonstrate that as \( J_{int} \) increases, the entire system gradually transitions from two single chains to a two-dimensional ladder and then back to a single chain, whereas the misaligned state cannot (the right side of Fig.\ref{fig13}). Above results indicate that when using dynamic indicators, not all initial states can accurately reflect the localization properties of the system, it is necessary to select appropriate initial states.

\subsection{\label{subsec4}Staggered configuration} 
\begin{figure*}[ht]	
  \centering
		\includegraphics[width=2\columnwidth]{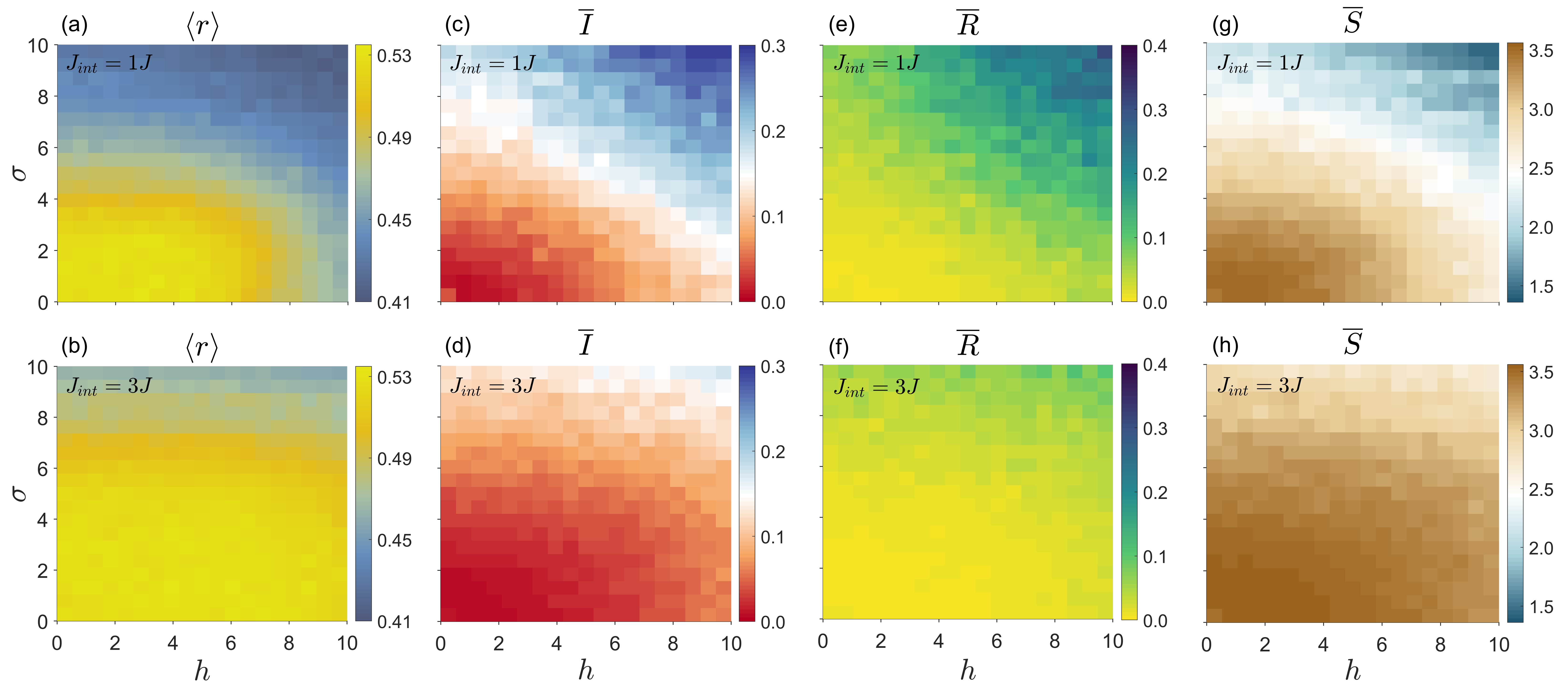}
		\caption{In the staggered configuration, the mean gap ratio $\langle r\rangle$ (first column, selecting 30 states near an energy density of $\varepsilon\approx0.5$ as samples), the steady-state staggered magnetization \(\overline{I}\) (second column), the steady-state return probability \(\overline{R}\) (third column), and the steady-state entanglement entropy \(\overline{S}\) (fourth column) of the entire system vary with Gaussian disorder $ \sigma $ and conventional disorder \( h \), for different values of \( J_{int} \). The upper four plots, $ J_{int} = 1J$. The bottom four plots, $ J_{int} = 10J$. Each point is averaged 100 disorder realisations.}
		\label{fig14}
	\end{figure*} 

  Next, we simulate cases where the coupling between the two parts of the entire system is more complex. We follow the same analytical approach as for the ladder configuration. First, we analyze the impact of the disorder from two parts. Fixing $ J_{int} = 1J$ and $ J_{int} = 3J$, we calculate the mean gap ratio, the steady-state staggered magnetization, the steady-state return probability, and the steady-state entanglement entropy as functions of the conventional disorder \(h\) of the upper chain and Gaussian disorder $ \sigma $ of the lower chain (Fig.\ref{fig14}). For the last three dynamic indicators, we choose the initial aligned state. From the results of the mean gap ratio (Fig.\ref{fig14}\textcolor{blue}{(a) (b)}), as the strengths of the disorder of two chains increase, the mean gap ratio gradually decreases, indicating that the system undergoes a localization phase transition. The other three dynamical steady-state indicators also yield consistent results (Fig.\ref{fig14}\textcolor{blue}{(c)-(h)}). The localization of the entire system is still jointly influenced by the disorder in both parts. The asymmetry caused by the different types of disorder remains present.

\begin{figure}[t]	
  \centering
		\includegraphics[width=1\columnwidth]{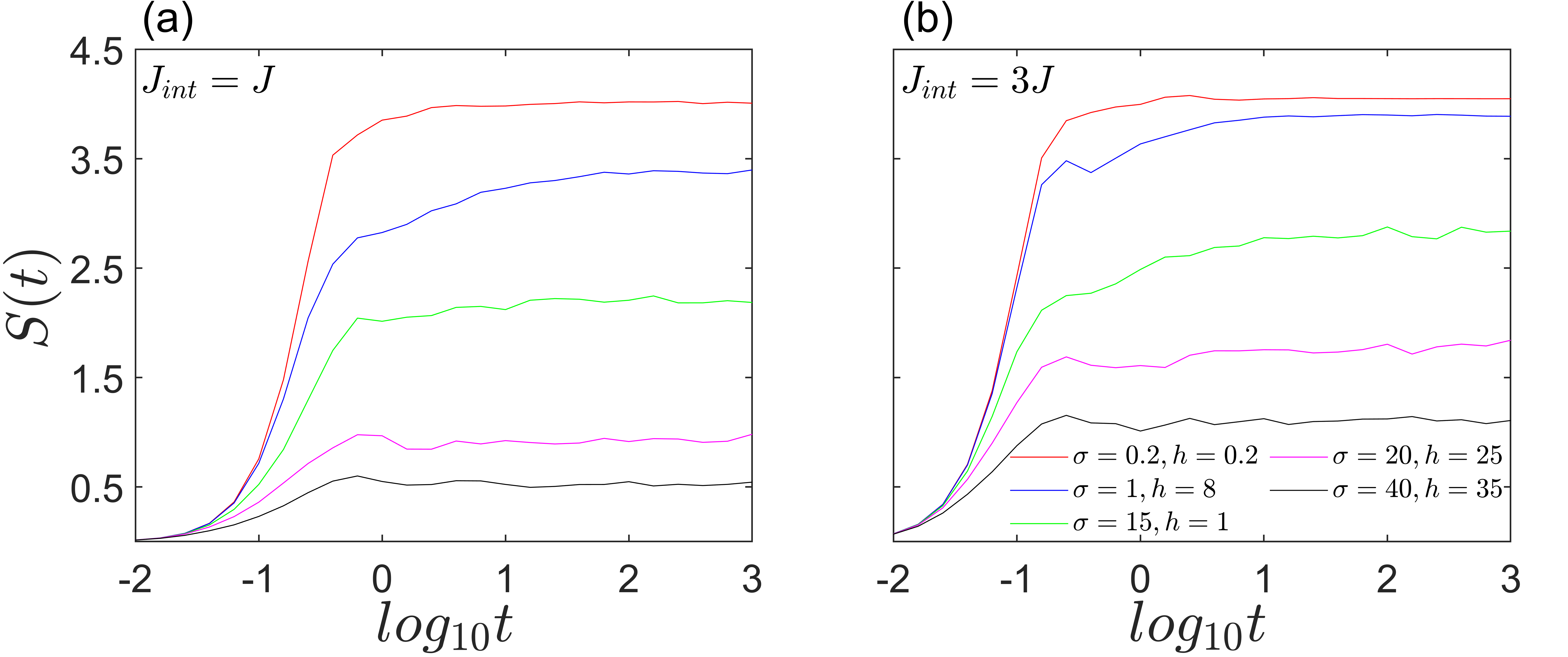}
		\caption{In the staggered configuration, the growth of entanglement entropy \(S(t)\) for different combinations of the Gaussian disorder $ \sigma $ and conventional disorder \( h \) with (a) $ J_{int} = 1J$ and (b) $ J_{int} = 3J$. Each point is averaged 1000 disorder realisations.}
		\label{fig15}
	\end{figure}

  Additionally, we use the evolution of entanglement entropy to distinguish between the localized and the ergodic phases. Based on the phase diagram of the mean gap ratio, we appropriately combine and select the disorder of two chains, and compute the evolution of entanglement entropy under two different coupling strengths (Fig.\ref{fig15}). It also shows that, for the weak disorder, the entanglement entropy grows logarithmically with time, while for the strong disorder, it grows faster with time.

\begin{figure}[t]	
  \centering
		\includegraphics[width=1\columnwidth]{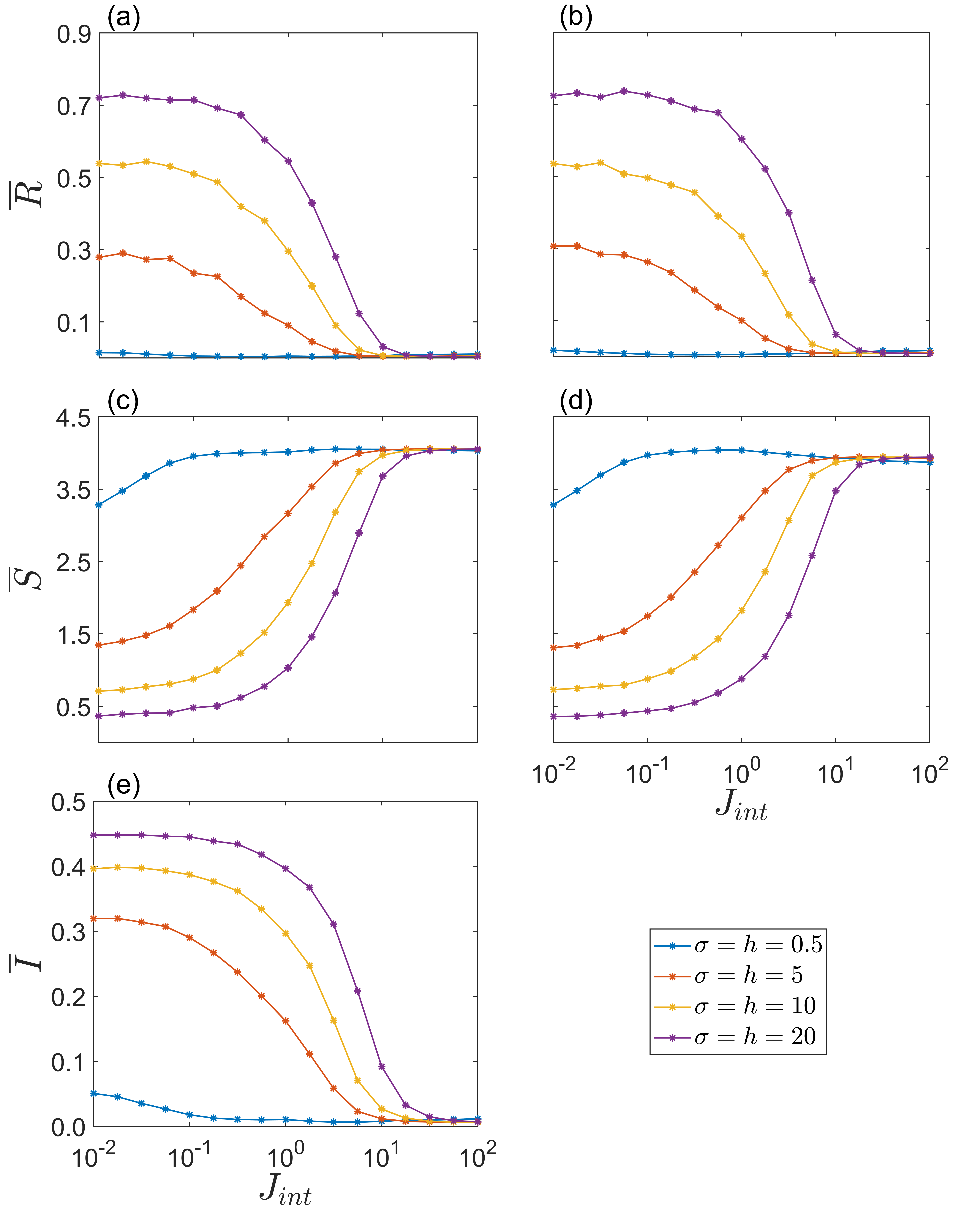}
		\caption{In the staggered configuration, (a) (b) the steady-state return probability \(\overline{R}\), (c) (d) the steady-state entanglement entropy \(\overline{S}\), and (e) the steady-state staggered magnetization \(\overline{I}\) vary with \( J_{int} \) for different values of disorder. The three plots on the left are for the initial aligned state. The two plots on the right are for the initial misaligned state. Each point is averaged 1000 disorder realisations.}
		\label{fig16}
	\end{figure}

  We also analyze the effect of \( J_{int} \) on the dynamical characteristics of the entire system for different initial states. For the staggered configuration, applying the same research methods as in Sec.\ref{subsec3}, we compute the steady-state return probability (Fig.\ref{fig16}\textcolor{blue}{(a) (b)}) and the steady-state entanglement entropy (Fig.\ref{fig16}\textcolor{blue}{(c) (d)}) for both initial aligned and misaligned states. The steady-state staggered magnetization (Fig.\ref{fig16}\textcolor{blue}{(e)}) is calculated only for the initial aligned state. $ \sigma=h=10$ and $ \sigma =h=0.5$ in Fig.\ref{fig16} correspond to the area above in Fig.\ref{fig9} ($\sigma =10$) and the area below in Fig.\ref{fig10} ($\sigma =0.5$), respectively. It shows that for both aligned and misaligned states, as \( J_{int} \) changes, the trends of all the indicators are the same, exhibiting a trend towards thermalization,  this is also consistent with the results in Sec.\ref{subsec2}.

  The results can be understood that as \( J_{int} \) increases, the entire system gradually transitions from two independent chains to a two-dimensional complex coupled model, which requires greater disorder to induce the MBL transition. Furthermore, as \( J_{int} \) continues to increase, the internal interaction strength within the complex coupled two-dimensional system further intensifies, necessitating an even larger disorder to achieve localization. Thus, for a fixed disorder strength, the indicators continuously trend towards thermalization as \( J_{int} \) increases.

  Regardless of whether the state is aligned or misaligned, the results of the three dynamic indicators are consistent. Additionally, as shown in Fig.\ref{fig14}, the dynamic indicators (Fig.\ref{fig14}\textcolor{blue}{(c)-(h)}) for the aligned initial state are consistent with the results of the mean gap ratio (Fig.\ref{fig14}\textcolor{blue}{(a)} and \textcolor{blue}{(b)}), which accurately reflects the localization properties of the many-body system. Therefore, for this configuration, both the aligned and misaligned states can accurately reflect the localization properties of entire system. So the dynamical results of these two configurations indicate that the dependence on the initial state is related with the structure of the entire system. For the staggered configuration, this initial-state dependence can be ignored.

\section{\label{sec6} Conclusion}
  The paper introduces a new type of disorder: Gaussian disorder. Gaussian disorder involves two parameters: the expected value $\mu$ and the standard deviation $\sigma$. Applying Gaussian disorder to a one-dimensional spin-1/2 chain, we find that the parameter driving the many-body localization phase transition is the standard deviation $\sigma$, while the parameter $\mu$ has no effect. Subsequently, with $\mu=0$, we study the properties of eigenstates and dynamical characteristics of the one-dimensional Gaussian disorder system. The results indicate that Gaussian disorder can induce the MBL phase transition, and as the Gaussian disorder $\sigma$ increases, the system gradually transitions from the ergodic phase to the localized phase. According to the characteristics of the Gaussian distribution, $\mu$ describes the central value of the disorder, while $\sigma$ describes the range of the disorder. Therefore, we can conclude that the factor influencing the localization properties of the system is the breadth of the disorder, not its central value.

  Additionally, we apply Gaussian disorder to the sys-env models, focusing on analyzing the impact they have on each other when the system and the environment are coupled. We use the steady-state staggered magnetization to indicate the eventual localization properties of the system with conventional disorder and the environment with Gaussian disorder. The coupling configuration, system disorder, environment disorder, interaction strength between system and environment, and initial state all affect their retention of initial state information. We place the system in localized and ergodic phases, respectively, and then compute how \(\overline{I}_{sys}\) and \(\overline{I}_{env}\) vary with the Gaussian disorder $ \sigma $ of the environment and the sys-env interaction strength \( J_{int} \) for different initial states. The results show that whether in the ladder or the staggered configuration, when $J_{i n t}\ll J$, the sys-env model simplifies into two independent chains. The system and environment maintain their initial localization properties due to their own disorder and are nearly unaffected by the disorder of the other. The system and environment have minimal impact on each other. When $J_{i n t}\approx J$, the two chains have started to establish a structural connection and are beginning to be influenced by each other's disorder. However, the dominant effect is still due to their own disorder strength. As \( J_{int} \) further increases, the system and environment show strong correlation, with ${\overline{I}}_{sys}\approx{\overline{I}}_{env}$. For the ladder configuration, the influence of environment (system) disorder on the system (environment) becomes more pronounced, with this effect being more evident in the ergodic system (environment). In addition, the results vary depending on the initial state. For the staggered configuration, whether the initial state is aligned or misaligned, both the system and environment completely lose their initial state information. This suggests that to enhance or reduce the retention of initial state information in a system coupled with the environment or in a subsystem, one can achieve this by adjusting the initial state or the coupling structure.

  Finally, we consider the system and environment as a whole to study the localization properties of the entire system. The results show that when the interaction strength \( J_{int} \) between two chains is fixed, as the disorder strength of each chain increases, the entire system gradually transitions from the ergodic phase to the localized phase. Moreover, even if the disorder strength of one chain is small, the entire system can still undergo the MBL transition, if the disorder strength of the other chain is sufficiently large. Then we fix the disorder strength and analyze the dynamical characteristics of the entire system change with \( J_{int} \) for different initial states. The results show that for the ladder configuration, the outcomes vary for different initial states (particularly noticeable at strong \( J_{int} \)), whereas for the staggered configuration, the dynamical results for different initial states are generally consistent. Because, in certain structured systems (such as the model of a strongly coupled ladder configuration), differences in initial states may lead to significant variations in energy density, which can lead to different localization properties. This means that for certain systems, we cannot determine the localization properties of the system solely based on the dynamical results from specific initial states. Actually, the dynamical results for the initial state only reflect the degree to which the system retains information of this initial state. It also indicates that a localized system cannot retain information for all initial states. Additionally, for certain models, not all initial states can accurately reflect the localization properties of the system. So if we want to reflect the system’s localization properties through its dynamical characteristics, we need to choose the initial state appropriately.

  \section{\label{sec7} Acknowledgement}
  This work was supported by the Plan for Scientific and Technological Development of Jilin Province (No.20230101018JC), by the NSF of China (Grant No.62175233), by the NSF of China (Grant No.62375259), and by the Plan for Scientific and Technological Development of Jilin Province (No.20220101111JC).

  \subsection{\label{A} Supplementary explanation that $\mu$ plays no role in triggering the MBL transition} 
 \begin{figure}[t]	
  \centering
		\includegraphics[width=1\columnwidth]{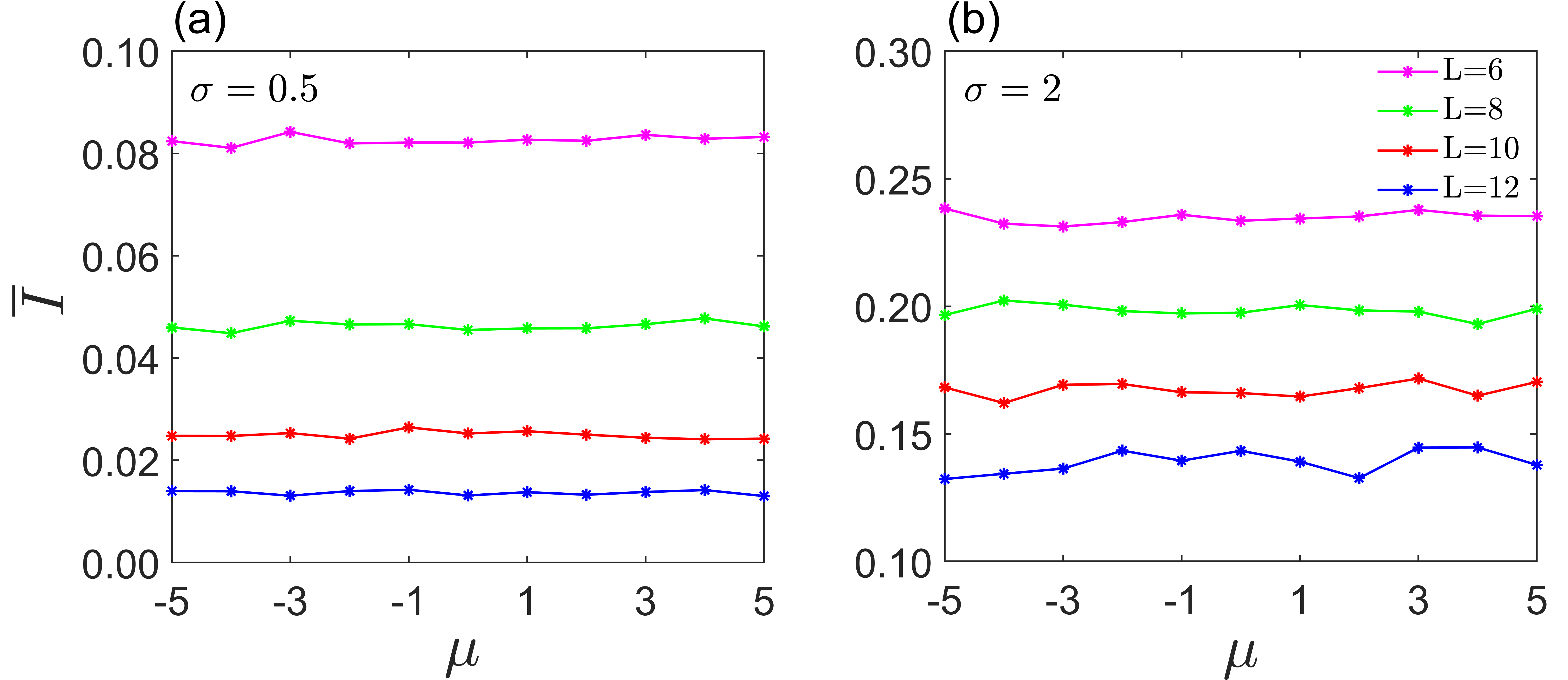}
		\caption{The steady-state staggered magnetization of the one-dimensional spin-1/2 chain for different chain lengths. (a) Gaussian disorder $ \sigma =0.5$. (b) Gaussian disorder $ \sigma =2$. Each point is averaged 1000 disorder realisations.}
		\label{fig17}
	\end{figure}

  In the Sec.\ref{sec3}, we demonstrate through von Neumann entropy that the parameter $\mu $ of Gaussian disorder does not influence the occurrence of the MBL transition. Here, based on the same model, we will further demonstrate this conclusion from a dynamical perspective using the steady-state staggered magnetization starting from the Néel state. Fig.\ref{fig17} present the variation of the steady-state staggered magnetization with $ \mu $ for different chain lengths. The results show that the steady-state staggered magnetization exhibits only minor fluctuations with changes in $ \mu $ . This also indicates that $ \mu $ does not influence the onset of the MBL transition. Therefore, the central value of disorder has no impact on the occurrence of the phase transition.

  \subsection{\label{B} Verification of translational invariance in conventional disorder} 

 \begin{figure}[t]
  \centering
		\includegraphics[width=1\columnwidth]{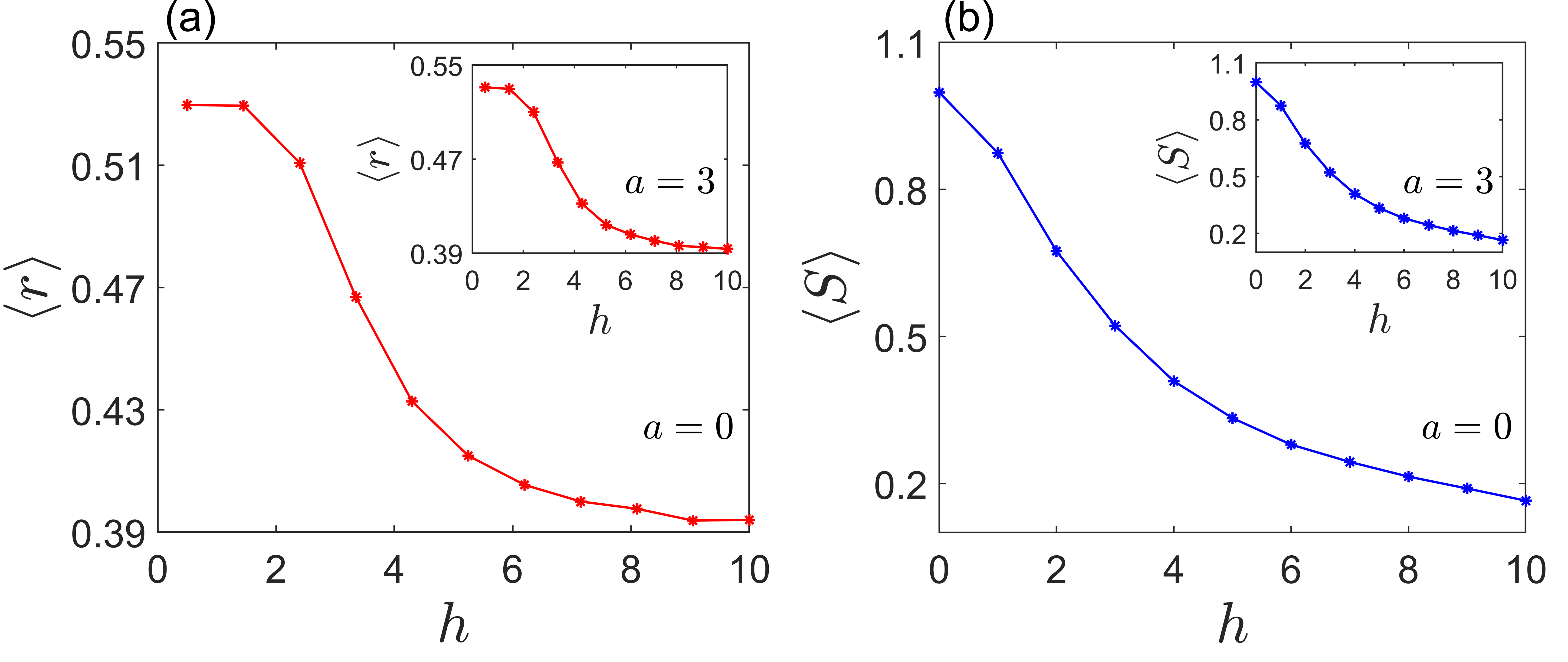}
		\caption{Consider a one-dimensional spin-1/2 chain for $L = 12$. Selecting 30 states near an energy density of $\varepsilon\approx0.5$ as samples. The variation of (a) the mean gap ratio $\langle r\rangle$ and (b) the entanglement entropy $\langle S\rangle$ with the change in disorder half-breadth \(h \). The main plot shows the central value $a = 0$, and the inset shows the central value $a = 3$. Each point is averaged 1000 disorder realisations.}
		\label{fig18}
        \end{figure}

  Based on the previous study of Gaussian disorder, we conclude that the factor influencing the occurrence of the phase transition is the breadth of the disorder, which is independent of the central value. Therefore, we infer that translating the conventional disorder to a larger or smaller interval of the same breadth has the same effect on MBL transition. That is, it is equivalent for the random distribution interval $\left[-h,h\right]$ (conventional disorder) and $\left[-h+a,h+a\right]$. Here, we use two physical quantities that can accurately reflect the properties of the system: the mean gap ratio $\langle r\rangle$ (Fig.\ref{fig18}\textcolor{blue}{(a)}) and the mean entanglement entropy $\langle S\rangle$ (Fig.\ref{fig18}\textcolor{blue}{(b)}) to verify our inference.

  We consider two cases with center values $a=0$ (Fig.\ref{fig18}) and 
  $a=3$ (the insets in Fig.\ref{fig18}). The former corresponds to the conventional disorder used in previous studies, while the latter corresponds to shifting the disorder interval of the former by 3 units in the positive direction. The results show that the outcomes are consistent in both cases, indicating that the central value \(a \) does not affect the MBL transition. Instead, it is the breadth of the disorder \(2h \), or simply \(h \), that influences the transition. This corresponds to the conclusions we obtained in Gaussian disorder.

  \subsection{\label{C}Another diagnostic of localization properties of the system and the environment} 
      
       \begin{figure}[t]
  \centering
		\includegraphics[width=1\columnwidth]{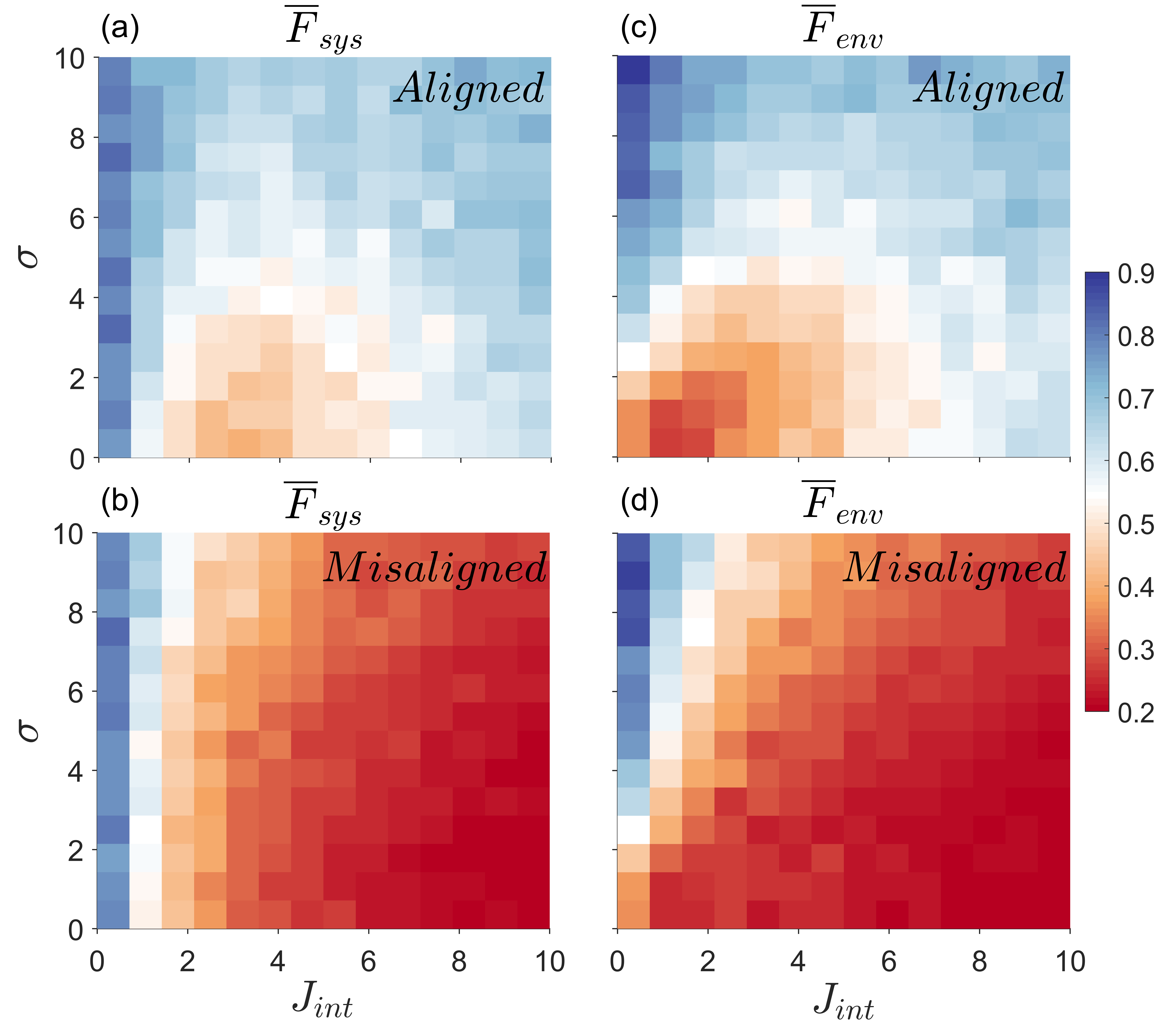}
		\caption{The conventional disorder of the system $h=10J$. The steady-state fidelity for the system \(\overline{F}_{sys}\) (left) and environment \(\overline{F}_{env}\) (right) for the ladder configuration. The upper two plots are for the initial aligned state. The bottom two plots are for the initial misaligned state. Each point is averaged 100 disorder realisations.}
		\label{fig19}
	\end{figure}

  Here we indicate the mutual influence between the system and the environment through another state-independent diagnostic. We use the fidelity instead of the staggered magnetisation as an indicator of localization. Fidelity ${\ F}(t)=\rm{Tr}\left[\rho(t)^{1/2}\rho(0)\rho(t)^{1/2}\right]^{1/2}$ can also be used to measure the deviation from an initial state\cite{zanardi2007information,zanardi2007mixed}. 
  
  We use ${\ F_{sys}}(t)=\rm{Tr}\left[\rho_{sys}(t)^{1/2}\rho_{sys}(0)\rho_{sys}(t)^{1/2}\right]^{1/2}$ and ${\ F_{env}}(t)=\rm{Tr}\left[\rho_{env}(t)^{1/2}\rho_{env}(0)\rho_{env}(t)^{1/2}\right]^{1/2}$ respectively to distinguish the eventual localization properties of the system and the environment. $\rho_{sys}\equiv\rm{Tr}_{env}\rho$ , $\rho_{env}\equiv\rm{Tr}_{sys}\rho$ and $\rho$ is chosen to be the pure state of the entire system composed of the system and the environment. Similarly, using the steady-state fidelity of the system ${\overline{{F}}}_{sys}$ and the environment ${\overline{{F}}}_{env}$ to indicate the extent to which the system and the environment retain their respective initial information. We take the ladder configuration, with the system initially in the localized phase at $h = 10J $, as an example. Fig.\ref{fig19} shows how the steady-state fidelity of the system and environment vary with the Gaussian disorder $ \sigma $ of the environment and the interaction strength \(J_{int} \) for different initial states.

  The results here are consistent with Fig.\ref{fig7} in Sec.\ref{sec4}. Therefore, the fidelity can replace the staggered magnetisation to analyze the mutual effects between the system and the environment.

\nocite{*}

\bibliography{apssamp}

\end{document}